\documentclass[a4paper,11pt]{article}

\pdfoutput=1
\usepackage{jheppub}
\usepackage{mathrsfs}
\usepackage{verbatim}
\usepackage{booktabs}
\usepackage[tight]{subfigure}

\usepackage[latin1]{inputenc} 
\usepackage{latexsym} 
\usepackage{graphicx} 
\usepackage{amsfonts}
\usepackage{amsmath} 
\usepackage{amssymb} 
\usepackage{parskip}
\usepackage{array}
\usepackage{color}

\usepackage{slashed}
\usepackage{epsfig,graphicx,bm,color} % Include figure files
\usepackage{dcolumn}% Align table columns on decimal point
\usepackage{bm}% bold math
\usepackage{hyperref}% add hypertext capabilities
\usepackage{breakurl}
\usepackage{soul} 
\newcommand{\be}{\begin{equation}}
\newcommand{\ee}{\end{equation}}
\newcommand{\bea}{\begin{eqnarray}}
\newcommand{\eea}{\end{eqnarray}}

\newcommand{\lsim}{\mathrel{\mathop{\kern 0pt \rlap
  {\raise.2ex\hbox{$<$}}}
  \lower.9ex\hbox{\kern-.190em $\sim$}}}
\newcommand{\gsim}{\mathrel{\mathop{\kern 0pt \rlap
  {\raise.2ex\hbox{$>$}}}
  \lower.9ex\hbox{\kern-.190em $\sim$}}}

\newcommand{\fb}{\mbox{ fb}}

\title{Model-independent combination of diphoton constraints at 750 GeV}

\author[a]{Jong Soo Kim,}
\emailAdd{jong.kim@csic.es}
\affiliation[a]{Instituto de F\'{\i}sica Te\'{o}rica, IFT-UAM/CSIC,\\
C/ Nicol\'{a}s Cabrera, 13-15, Cantoblanco, ES-28049 Madrid, Spain}

\author[a,b]{Krzysztof Rolbiecki,}
\emailAdd{krzysztof.rolbiecki@desy.de}
\affiliation[b]{Institute of Theoretical Physics, University of Warsaw, PL-02093 Warsaw, Poland}

\author[c]{Roberto Ruiz de Austri}
\emailAdd{rruiz@ific.uv.es}
\affiliation[c]{Instituto de F\'isica Corpuscular, IFIC-UV/CSIC, Valencia, Spain}

\abstract{
Motivated by the recent diphoton excess reported by both the ATLAS and
CMS collaborations, we provide a model-independent combination of diphoton results obtained at $\sqrt{s}=8$ and 13~TeV at the LHC. We consider resonant $s$-channel production of a spin-0 and spin-2 particle with a mass of $750$~GeV that subsequently 
decays to two photons. The size of the excess reported by ATLAS appears to be in a slight tension with other measurements under the spin-2 particle hypothesis.}  


\preprint{IFT-UAM/CSIC-15-139}
\begin{document}
\maketitle
\flushbottom

%%%%%%%%%%%%%%%%%%%%%%%%%%%%%%%%%%%%%%%%
%%%%%%%%%%%%%%%%%%%%%%%%%%%%%%%%%%%%%%%%

\section{Introduction}

The ATLAS and CMS collaboration have found an excess in the search for a diphoton final state after 
the first $13$ TeV data have been analyzed \cite{ATLASdiphoton2015,CMSdiphoton2015}. 
The excess points to a resonance with an invariant mass of about 750~GeV with a local significance 
of $3.6 \, \sigma$ (ATLAS) and $2.6 \, \sigma$ (CMS).

The simplest explanation of the excess is through resonant production of a
spin-0 or spin-2 particle with a mass of around $750$ GeV that decays to photons. A spin-1 
resonance is excluded by the Yang--Landau theorem~\cite{Landau:1948kw,Yang:1950rg}.
There have been many attempts to explain the excess both via direct production of the 750~GeV resonance or
through a heavier particle that decays on-shell to a pair of $750$~GeV
scalars finally decaying to photons \cite{Harigaya:2015ezk,Mambrini:2015wyu,Backovic:2015fnp,Angelescu:2015uiz,
Buttazzo:2015txu,Knapen:2015dap,Nakai:2015ptz,Pilaftsis:2015ycr,Franceschini:2015kwy,DiChiara:2015vdm,Higaki:2015jag,
McDermott:2015sck,Ellis:2015oso,Low:2015qep,Bellazzini:2015nxw,Gupta:2015zzs,Petersson:2015mkr,Molinaro:2015cwg,
Dutta:2015wqh,Cao:2015pto,Matsuzaki:2015che,Kobakhidze:2015ldh,Martinez:2015kmn,Cox:2015ckc,Becirevic:2015fmu,
No:2015bsn,Demidov:2015zqn,Chao:2015ttq,Fichet:2015vvy,Curtin:2015jcv,Bian:2015kjt,Chakrabortty:2015hff,Ahmed:2015uqt,
Agrawal:2015dbf,Csaki:2015vek,Falkowski:2015swt,Aloni:2015mxa,Bai:2015nbs,Gabrielli:2015dhk,Benbrik:2015fyz,
Kim:2015ron,Alves:2015jgx,Megias:2015ory,Carpenter:2015ucu,Bernon:2015abk,Buckley:2016mbr}; see Ref.~\cite{Staub:2016dxq} for a recent review.

In this letter we investigate whether the interpretation of the diphoton excess via 
resonant $s$-channel production is compatible with the full set of Run-I data~\cite{Aad:2014ioa,Aad:2015mna,Khachatryan:2015qba} for both the spin-0 and the spin-2 particle hypotheses.
We work in a model-independent framework in which we parametrize the diphoton rate by the cross section and branching ratio to photons 
and perform a simple statistical test to assess the compatibility between different measurements.

This work is structured as follows. In Section~\ref{sec:meth} we explain the methodology that we have employed, in Section~\ref{sec:results} we present the results and finally we give our conclusions in the last section.

%%%%%%%%%%%%%%%%%%%%%%%%%%%%%%%%%%%%%%%%%%%%%%%%
%%%%%%%%%%%%%%%%%%%%%%%%%%%%%%%%%%%%%%%%%%%%%%%%

\section{Methodology\label{sec:meth}}
We assume that the diphoton signal is resonantly produced,
\begin{equation}
pp\rightarrow X\rightarrow\gamma\gamma\,, 
\end{equation}
where $X$ denotes either a spin-0 or spin-2 particle.
Here, we consider the case where the resonance is only produced via gluon fusion~\cite{Barger:1987nn},
\begin{equation}
\sigma(pp\rightarrow X)=(2J+1)\Gamma(X\rightarrow gg) \frac{\pi^2}{8 m_X^3} \tau \int_{\tau}^{1}\frac{dx}{x} g\left(x,m_X^2\right) g\left(\tau,m_X^2\right)\,,
\end{equation}
where we have introduced the dimensionless variable $\tau=\frac{m_X^2}{s}$. $J$ and $g(x,m_X^2)$ denotes the spin of the resonance and the gluon distribution function of the proton, respectively. 
Note that the gluon luminosity ratio between $13$ and $8$~TeV is $4.7$ for $m_X=750$~GeV~\cite{Higgs:Cross}.
The branching ratio into the diphoton final state is given by
\begin{equation}
\mathrm{BR}(X\rightarrow \gamma\gamma)= \frac{\Gamma(X\rightarrow\gamma\gamma)}{\Gamma(X\rightarrow\gamma\gamma) + \Gamma(X\rightarrow gg) + \Gamma(X\rightarrow YY)}\,,
\end{equation}
where $Y$ denotes all other particles which can couple to the resonance $X$. Due to the much lower increase in the $u$ and $d$ quarks luminosity between $\sqrt{s} = 8$ and $13$~TeV of $2.5$--$2.7$~\cite{Franceschini:2015kwy}, the production in quark--antiquark annihilation would lead to significant tensions between $8$ and $13$~TeV results as we will see later. For this reason we will ignore this possibility in the following. This is different for heavy quark initial states: the cross section increase for producing a 750 GeV resonance is $5.1$--$5.4$ for charm and bottom initial states, hence numerically close to the enhancement in gluon--gluon production. Therefore, our results would be qualitatively valid also in this case, albeit with a reduced tension, for a detailed analysis see Ref.~\cite{Domingo:2016unq}.\footnote{It was pointed out in Ref.~\cite{Bernon:2016dow} that the gluon--gluon and $b\bar{b}$-initiated production could be experimentally distinguishable by looking at $p_T$ distributions of photons and jets as well as a number of additional jets.}

A sample of signal events for the spin-0 case was generated with {\tt POWHEG}~\cite{Nason:2004rx,Frixione:2007vw,Alioli:2010xd} 
at the parton level and interfaced with {\tt Pythia 6.4}~\cite{Sjostrand:2006za} for the parton shower and hadronization with the {\tt CTEQ6L1} parton distribution function~\cite{Lai:2010vv}. A sample for the spin-2 case was generated with {\tt Herwig++ 2.7.1}~\cite{Bahr:2008pv} using the MSTW parton distribution functions \cite{Martin:2009iq}. For both hypotheses we assume a decay width of 45~GeV.
We have implemented the $8$~TeV~\cite{Aad:2015mna,Aad:2014ioa,Khachatryan:2015qba,CMS-PAS-EXO-12-045} and $13$ TeV~\cite{ATLASdiphoton2015,CMSdiphoton2015} diphoton searches from ATLAS and CMS
into the {\tt CheckMATE 1.2.2} framework~\cite{Drees:2013wra} with its {\tt AnalysisManager}~\cite{Kim:2015wza}. 
{\tt CheckMATE 1.2.2} is based on the fast detector simulation {\tt Delphes 3.10}~\cite{deFavereau:2013fsa} with heavily 
modified detector tunes and it determines the number of expected signal events passing the selection cuts of the particular analysis. 
The cuts of the ATLAS and CMS analyses are shown in Table~\ref{tab:selection13tev}.
We do not follow the approach of both experiments where the expected signal plus background distribution is fitted to the measured $m_{\gamma\gamma}$ distribution. Instead, we just perform a simple cut-and-count study. Our simplified implementation of the analysis certainly leads to a reduced sensitivity, however, our conclusions will still be viable and can be regarded as more conservative. As a result, the signal regions of all employed searches are defined as $700 < m_{\gamma\gamma} < 800$~GeV, except for the ATLAS exotic search~\cite{Aad:2015mna} where we use the original signal region with
$719 < m_{\gamma\gamma} < 805$~GeV. Since the size of the mass bin in the signal regions are relatively large, our conclusions do not depend on the exact value of the total decay width as long as we do not assume a very broad resonance.
\begin{table}
\centering \renewcommand{\arraystretch}{1.3}
\begin{tabular}{|c|c|}\hline
 ATLAS & CMS \\
\hline
\hline
$p_T(\gamma)\ge$25 GeV  &
$p_T(\gamma)\ge$75 GeV \\ 
\hline
$|\eta^{\gamma}|\le2.37$ & $|\eta^{\gamma}|\le 1.44$
 or $1.57 \le |\eta^{\gamma}| \le 2.5$ 
\\ 
& at least one $\gamma$ with $|\eta^{\gamma}|\le1.44$\\
\hline
$E_T^{\gamma_1}/m_{\gamma\gamma}\ge0.4$,
$E_T^{\gamma_2}/m_{\gamma\gamma}\ge0.3$ & $m_{\gamma\gamma}\ge230$ GeV \\ \hline
\end{tabular}
\caption{Selection cuts of the 13 TeV ATLAS/CMS diphoton searches
  \cite{ATLASdiphoton2015,CMSdiphoton2015}. 
\label{tab:selection13tev} }
\end{table}

In order to test the implementation of the analyses within the {\tt CheckMATE} framework we performed a number of tests. For all searches, the acceptance times efficiency is typically provided in the publications and this can be compared to our Monte Carlo (MC) simulation. In Table~\ref{tab:validation} we compare our efficiency with the efficiency computed by the experimental collaboration. In addition to the efficiency numbers, we provide additional information for each channel. The relative differences between the efficiencies obtained by {\tt CheckMATE} and the one reported by the experiments are typically around $10\%$. 

\begin{table}
\begin{center}
\renewcommand{\arraystretch}{1.3} 
\scalebox{0.8}{
\begin{tabular}{|l||c|c||c|c|}\hline
 search & exp.\ eff. & comments &  {\tt CheckMATE} & comments  \\
\hline
CMS13 EBEB~\cite{CMSdiphoton2015} & $34\%$ & RS graviton, $m=750$~GeV & $38\%$ & RS graviton, $m=750$~GeV \\
CMS13 EBEE~\cite{CMSdiphoton2015} & $22\%$ & RS graviton, $m=750$~GeV & $23\%$ & RS graviton, $m=750$~GeV \\
ATLAS13~\cite{ATLASdiphoton2015} & $>40\%$ & scalar, gluon fusion, $m>600$~GeV & $44\%$ & scalar $m=750$~GeV\\
ATLAS8 HIG~\cite{Aad:2014ioa} & $56$--$71\%$ & scalar & $57\%$ & scalar $m=750$~GeV\\
ATLAS8 EXO~\cite{Aad:2015mna} & $50\%$ & RS graviton, $m=2$~TeV & $55\%$ & RS graviton, $m=2$~TeV\\
CMS8 EXO~\cite{CMS-PAS-EXO-12-045} & $40\%$ & RS graviton, $m=750$~GeV & $45\%$ & RS graviton, $m=750$~GeV \\
CMS8 HIG~\cite{Khachatryan:2015qba} & $75\%$ & narrow width scalar $m=750$~GeV & $65\%$ & narrow width scalar $m=750$~GeV \\ \hline
\end{tabular}}
\caption{Validation of the {\tt CheckMATE} implementation of the ATLAS and CMS diphoton searches. 
\label{tab:validation} }
\end{center}
\end{table}

The goal is to perform a statistical test of the spin-0 or spin-2 hypothesis taking the
8 and 13~TeV data of the ATLAS and CMS experiment as input, separately as well as combined.
The fit was performed with the $\chi^2$ test statistics.\footnote{This is reasonable since the number 
of observed events in most of the signal regions is $> 20$.} Namely, 
\begin{eqnarray}\label{eq:chi2}
 \chi_i^2 = \frac{(n_i-\mu_i)^2}{\sigma^2_{i,{\rm stat}}+\sigma_{i,b}^2}\;, 
\end{eqnarray}
where
\begin{eqnarray}
 \mu_i = \mu_{i,b}+\mu_{i,s}\;.\nonumber
 \end{eqnarray}
Here, $n_i$ is the number of observed events, $\mu_{i,b}$ is the expected number of background 
events, $\mu_{i,s}$ is the expected number of signal 
events, $\sigma_{i,{\rm stat}}$ and $\sigma_{i,b}$ are the statistical and systematic uncertainty 
on the expected number of background events for each signal region, $i$. The total systematic errors combine the systematic errors given by the collaborations (c.f.\ Table~\ref{tab:result-spin2}) and a $10\%$ error on the
\texttt{CheckMATE} event yield. For the statistical error we assume that it follows the Poisson distribution. 

The fit is performed for three cases: using only ATLAS $13$ TeV result; using both measurements at $13$~TeV [$i = $ `ATLAS13', `CMS13 EBEB', `CMS13 EBEE' in Eq.~\eqref{eq:chi2}]; and finally using $13$~TeV results and exotic searches at $8$~TeV~\cite{Aad:2015mna,CMS-PAS-EXO-12-045} [$i = $ `ATLAS13', `CMS13 EBEB', `CMS13 EBEE', `ATLAS8 EXO', `CMS8 EXO' in Eq.~\eqref{eq:chi2}]. We do not combine searches from the same experiment at $8$~TeV as these are highly correlated. In the following, we will see that the ATLAS exotic search at $\sqrt{s}=8$~TeV has a better reach compared to the other 8~TeV diphoton searches. Namely, a potentially higher sensitivity and well defined signal regions which motivated our choice. 
We use the following conventions for different searches and signal regions:
ATLAS13~\cite{ATLASdiphoton2015}; CMS13~\cite{CMSdiphoton2015} EBEB, EBEE for the barrel end-cap signal regions; ATLAS8 HIG~\cite{Aad:2014ioa};  ATLAS8 EXO~\cite{Aad:2015mna}; 
CMS8 EXO~\cite{CMS-PAS-EXO-12-045}; and CMS8 HIG~\cite{Khachatryan:2015qba}.

%%%%%%%%%%%%%%%%%%%%%%%%%%%%%%%%%
%%%%%%%%%%%%%%%%%%%%%%%%%%%%%%%%%

\section{Results and discussion}\label{sec:results}

Our results are summarized in Table~\ref{tab:result-spin2} for the spin-2 resonance and in Table~\ref{tab:result-spin0} for a spin-0 boson. For
each signal region we give the observed number of events, the expected number of background events, its total error and the results for three different fits using different sets of measurements. For example, 
using the ATLAS13 measurement only, we expect $16.5$ signal events for the best-fit point solution which translates to $40.0$ signal events in the ATLAS8 EXO search corresponding to a CL$_{\mathrm{S}}$ value of $8.4\cdot10^{-4}$~\cite{Read:2002hq} which is clearly excluded (see columns 4 and 5 of Table~\ref{tab:result-spin2}). Already at this point we can say that the 8~TeV ATLAS result is in tension with other measurements under our model hypothesis. 

\begin{table*}[t!]
\begin{center}
\scalebox{0.89}{
\begin{tabular}{|c||c|c||c|c||c|c||c|c|}
\hline
signal region & observed & background & best fit & $\Delta \chi^2$ & best fit & $\Delta \chi^2$ & best fit & $\Delta \chi^2$ \\  %& Best Fit Point\\
\hline
& & & \multicolumn{2}{c||}{ATLAS13} & \multicolumn{2}{c||}{ATLAS13+CMS13} & \multicolumn{2}{c|}{combined}\\
\hline
ATLAS13    & $28$ & $11.4\pm 3$    & $16.5$ & $-$    & $6.7$  & $2.7$ & $3.9$ & $4.3$  \\
\hline
CMS13 EBEB & $14$ & $9.5 \pm 1.9$  & $16.5$ & $8.2$  & $6.7$  & $0.3$ & $3.9$ & $0.0$  \\
\hline
CMS13 EBEE & $16$ & $18.5 \pm 3.7$ & $10.2$ & $5.4$  & $4.1$  & $1.5$ & $2.4$ & $0.8$  \\
\hline
ATLAS8 HIG & $34$ & $28 \pm 5$     & $22.4$ & $4.5$  & $9.0$  & $0.2$ & $5.3$ & $0.0$  \\
\hline
ATLAS8 EXO & $99$ & $96.4\pm 3.2$  & $40.0$ & $12.8$ & $16.1$ & $1.7$ & $9.5$ & $0.4$  \\
\hline
CMS8 EXO   & $46$ & $48.6 \pm 5.4$ & $15.0$ & $4.0$  &  $6.1$ & $0.4$ & $3.6$ & $0.1$  \\
\hline
CMS8 HIG   & $53$ & $50 \pm 6$     & $22.0$ & $4.12$ &  $8.9$ & $1.0$ & $5.2$ & $0.5$  \\
\hline
\end{tabular}}
\end{center}
\caption{The number of events in each signal region for production of a spin-2 particle: observed, SM background, our 'best fit' according to the MC results and the $\Delta \chi^2$ contribution. Each simulated signal region is compared to three cases, where the best-fit point is obtained using: ATLAS13 data only; CMS13 and ATLAS13 combined; ATLAS and CMS from 8 and $13$ TeV combined; see text for details.\label{tab:result-spin2}}
\end{table*}

\begin{table*}[t!]
\begin{center}
\scalebox{0.89}{
\begin{tabular}{|c||c|c||c|c||c|c||c|c|}
\hline
signal region & observed & background & best fit & $\Delta \chi^2$ & best fit & $\Delta \chi^2$ & best fit & $\Delta \chi^2$ \\  %& Best Fit Point\\
\hline
& & & \multicolumn{2}{c||}{ATLAS13} & \multicolumn{2}{c||}{ATLAS13+CMS13} & \multicolumn{2}{c|}{combined}\\
\hline
ATLAS13    & $28$ & $11.4\pm 3$    & $16.6$ & $-$    & $9.6$  & $1.3$ & $6.6$  & $2.7$  \\
\hline
CMS13 EBEB & $14$ & $9.5 \pm 1.9$  & $14.4$ & $2.1$  & $8.3$  & $1.1$ & $5.8$  & $0.7$  \\
\hline
CMS13 EBEE & $16$ & $18.5 \pm 3.7$ & $5.4$  & $5.6$  & $3.1$  & $0.8$ & $2.2$  & $0.1$  \\
\hline
ATLAS8 HIG & $34$ & $28 \pm 5$     & $20.7$ & $3.7$  & $12.0$ & $0.6$ & $8.3$  & $0.1$  \\
\hline
ATLAS8 EXO & $99$ & $96.4\pm 3.2$  & $28.1$ & $5.9$  & $16.2$ & $1.7$ & $11.2$ & $0.7$  \\
\hline
CMS8 EXO   & $46$ & $48.6 \pm 5.4$ & $12.2$ & $3.7$  &  $7.1$ & $1.0$ & $4.9$  & $0.3$  \\
\hline
CMS8 HIG   & $53$ & $50 \pm 6$     & $21.1$ & $2.9$ &  $12.2$ & $1.2$ & $8.4$  & $0.7$  \\
\hline
\end{tabular}}
\end{center}
\caption{The number of events in each signal region for production of a spin-0 particle: observed, SM background, our 'best fit' according to the simulation results and the $\Delta \chi^2$ contribution. Each simulated signal region is compared to three cases, where the best-fit point is obtained using: ATLAS13 data only; CMS13 and ATLAS13 combined; ATLAS and CMS from 8 and $13$~TeV combined; see text for details. \label{tab:result-spin0} }
\end{table*}

\begin{figure}[t!]
\begin{center}                                                                                                                                  
\includegraphics[width=0.32\textwidth]{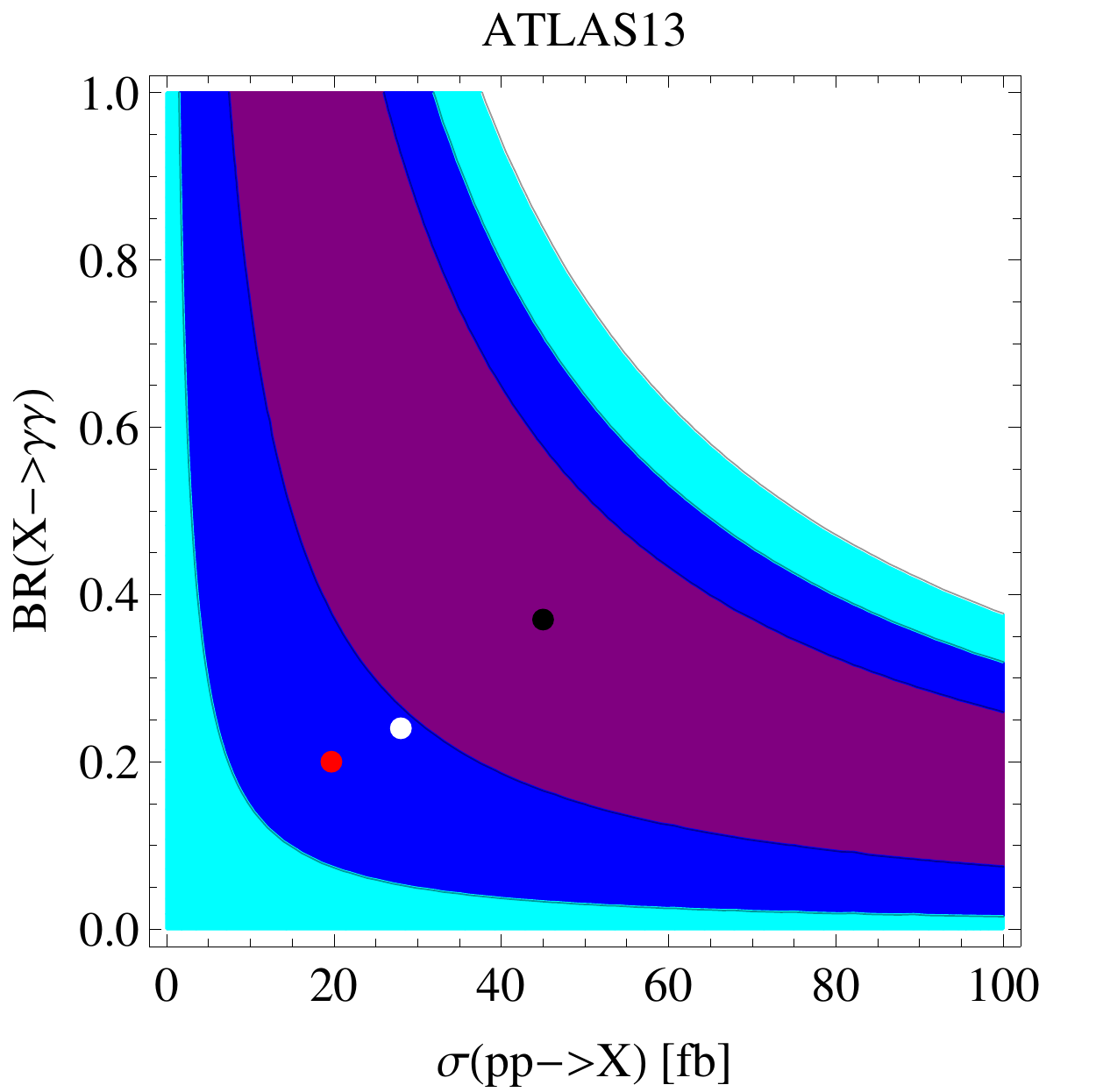} 
\includegraphics[width=0.32\textwidth]{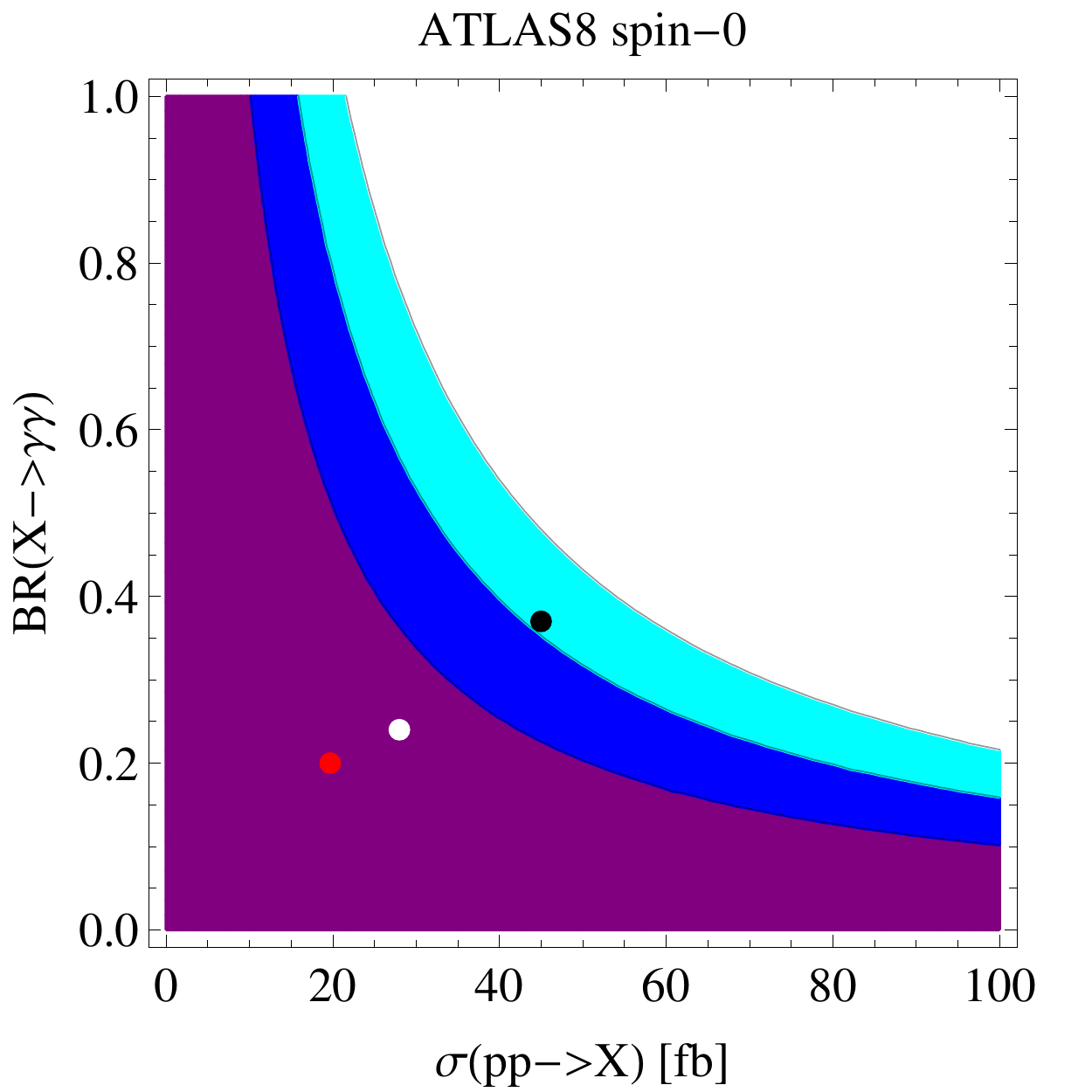}
\includegraphics[width=0.32\textwidth]{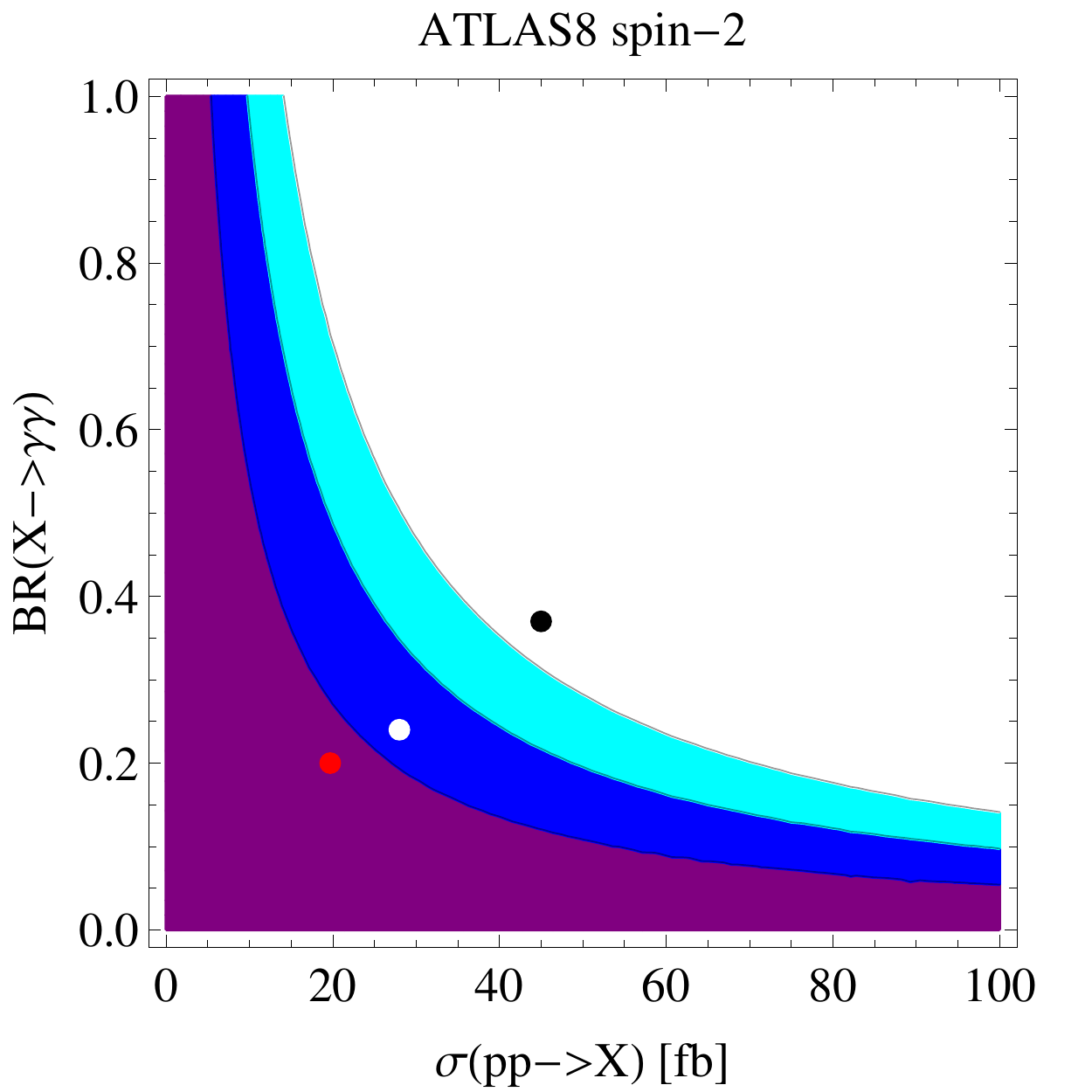}\\
%\vspace{-0.75cm}
\hspace{0.05\textwidth}(a) \hspace{0.28\textwidth} (b) \hspace{0.28\textwidth} (c)
\caption{The distribution of the $\chi^2$ test as a function of BR$(X\rightarrow\gamma\gamma)$ and $\sigma(pp \to X)$, where $X$ is a spin-2 particle, for (a) ATLAS13~\cite{ATLASdiphoton2015}; (b) ATLAS8 HIG~\cite{Aad:2014ioa}; and (c) ATLAS8 EXO~\cite{Aad:2015mna}. The colors denote: purple 1-$\sigma$ compatibility; blue 2-$\sigma$ compatibility; and cyan 3-$\sigma$. The dots represent sample best-fit points: the black point corresponds to the fit using only ATLAS13, the white point using both 13~TeV results and the red point for a combination of 8 and 13~TeV. \label{fig:atlas_2}}
\end{center}
\end{figure}

\begin{figure}[t!]
\begin{center}                                                                                                                                   
\includegraphics[width=0.32\textwidth]{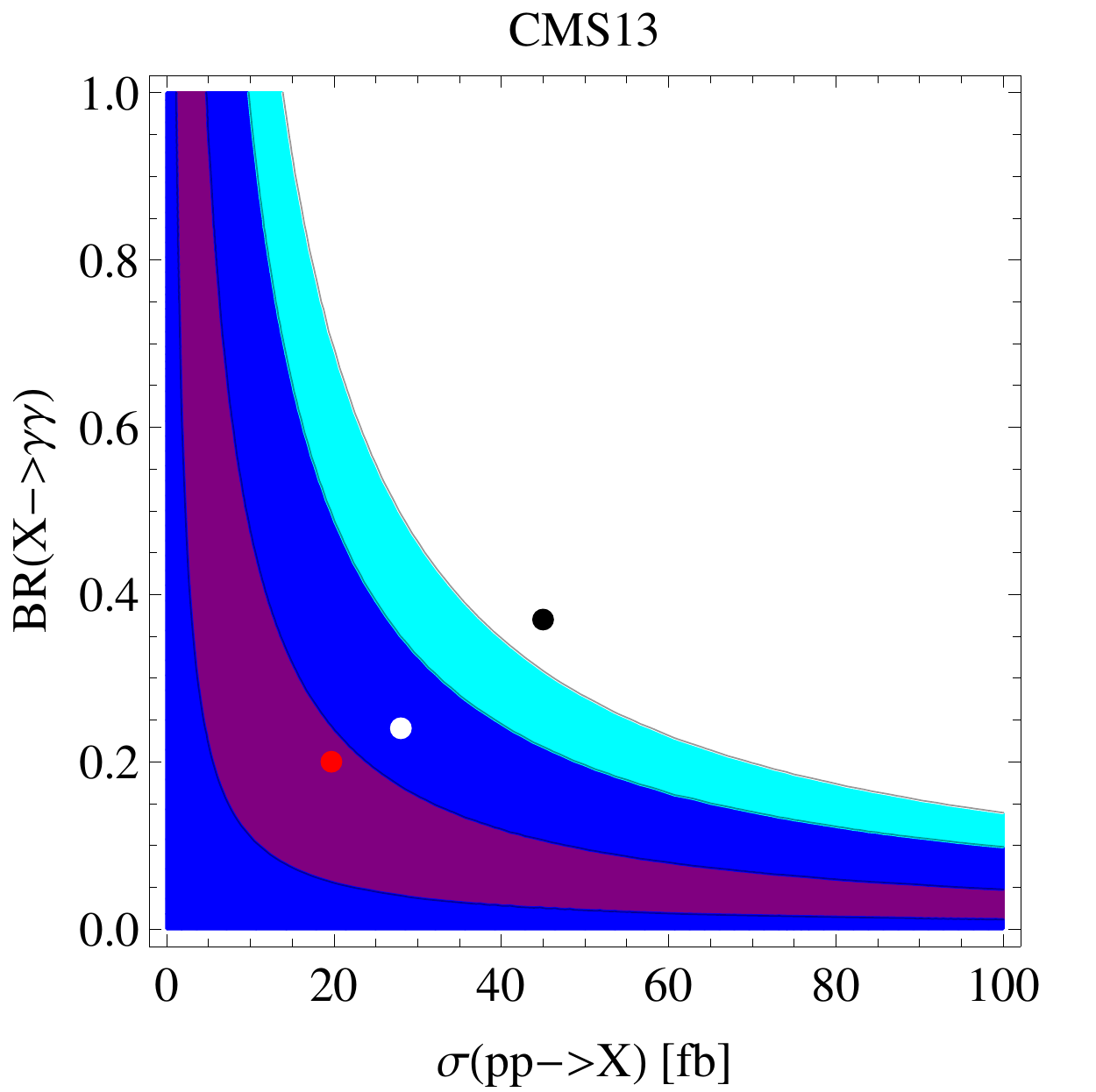} 
\includegraphics[width=0.32\textwidth]{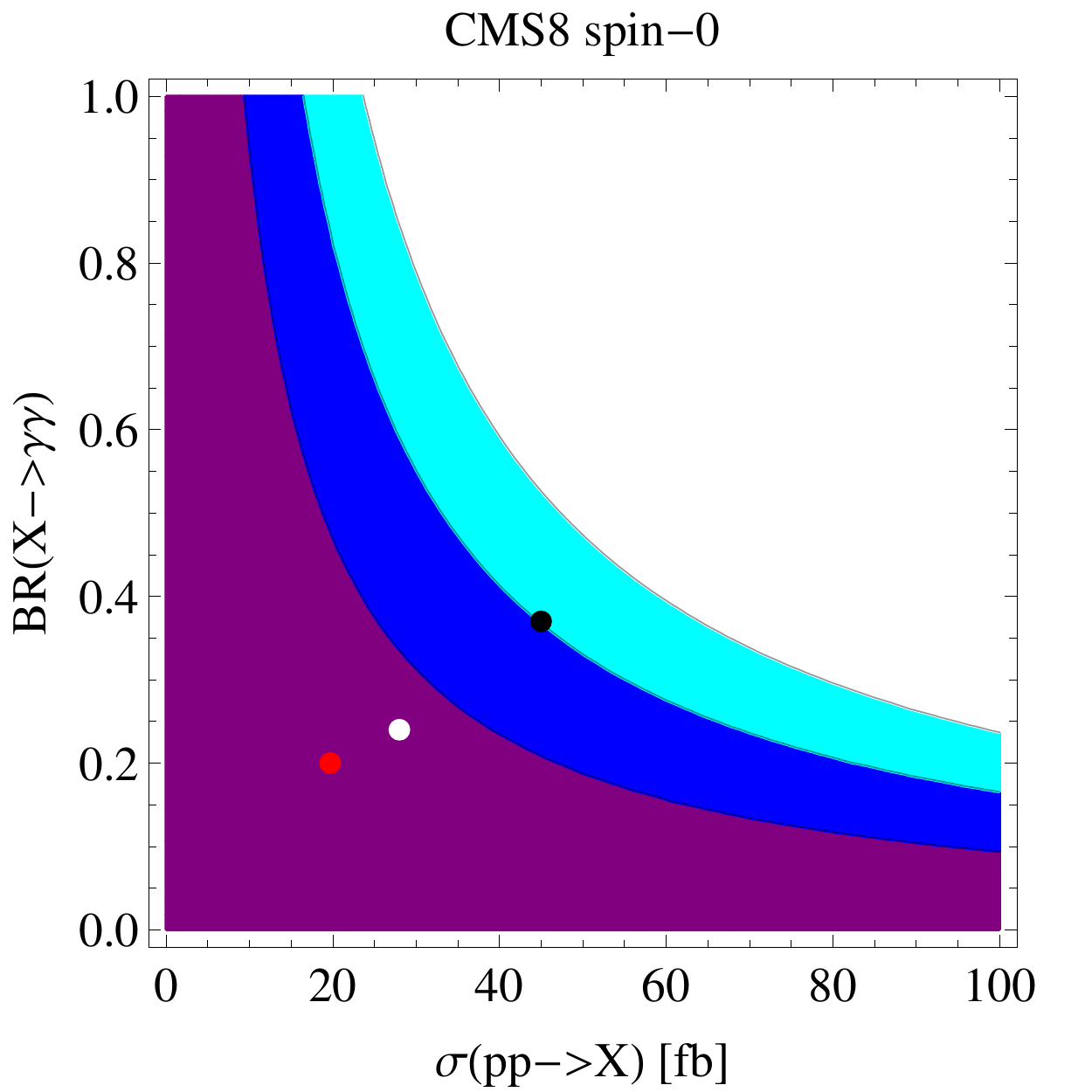} 
\includegraphics[width=0.32\textwidth]{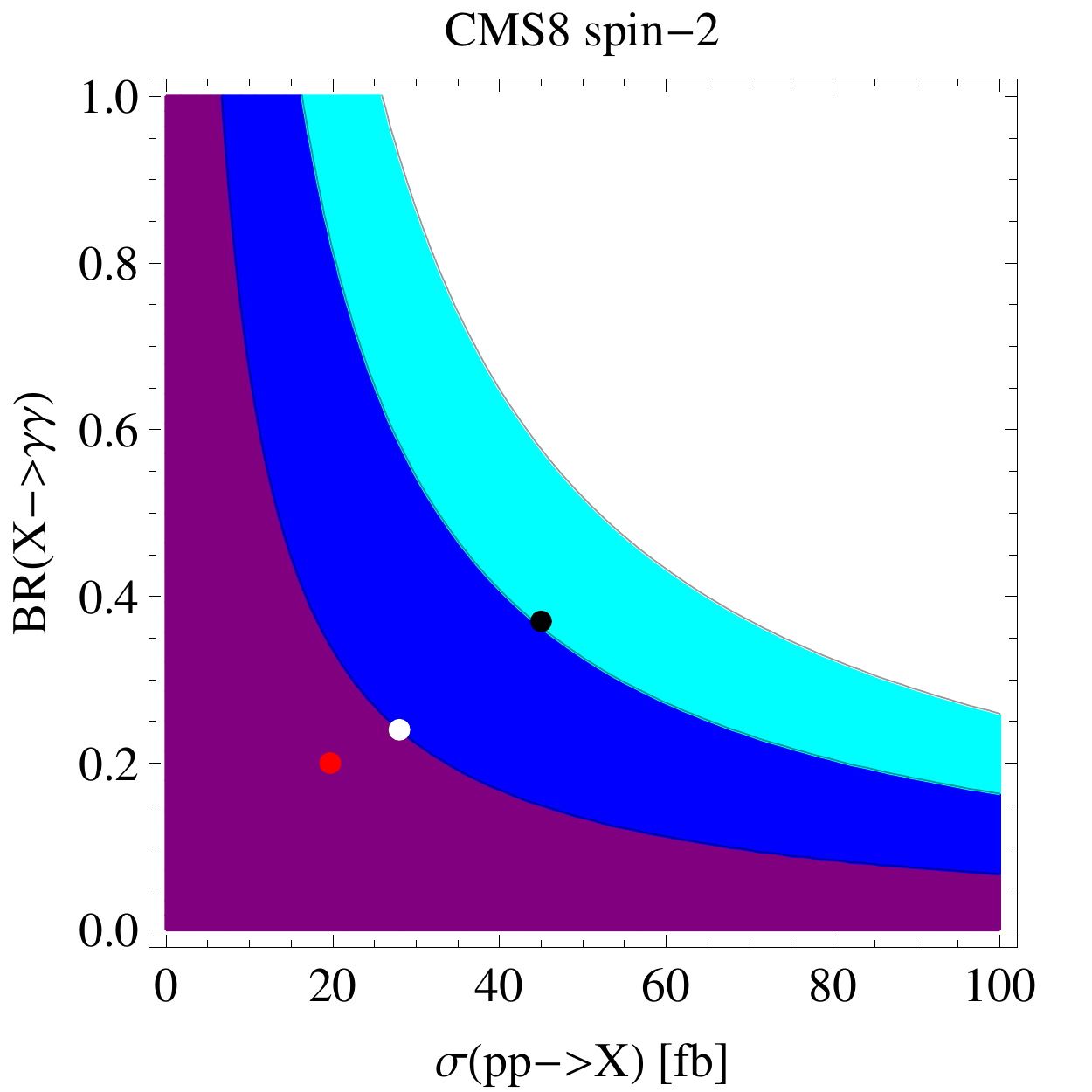} \\
%\vspace{-0.75cm}
\hspace{0.05\textwidth}(a) \hspace{0.28\textwidth} (b) \hspace{0.28\textwidth} (c)
\caption{The distribution of the $\chi^2$ test as a function of BR$(X\rightarrow\gamma\gamma)$ and $\sigma(pp \to X)$, where $X$ is a spin-2 particle, for (a) CMS13~\cite{CMSdiphoton2015}; (b) CMS8 HIG~\cite{Khachatryan:2015qba}; and (c) CMS8 EXO~\cite{CMS-PAS-EXO-12-045}. See text and Figure~\ref{fig:atlas_2} for more details. }
\label{fig:cms_2}
\end{center}
\end{figure}

\begin{figure}[t!]
\begin{center}                                                                                                                                   
\includegraphics[width=0.45\textwidth]{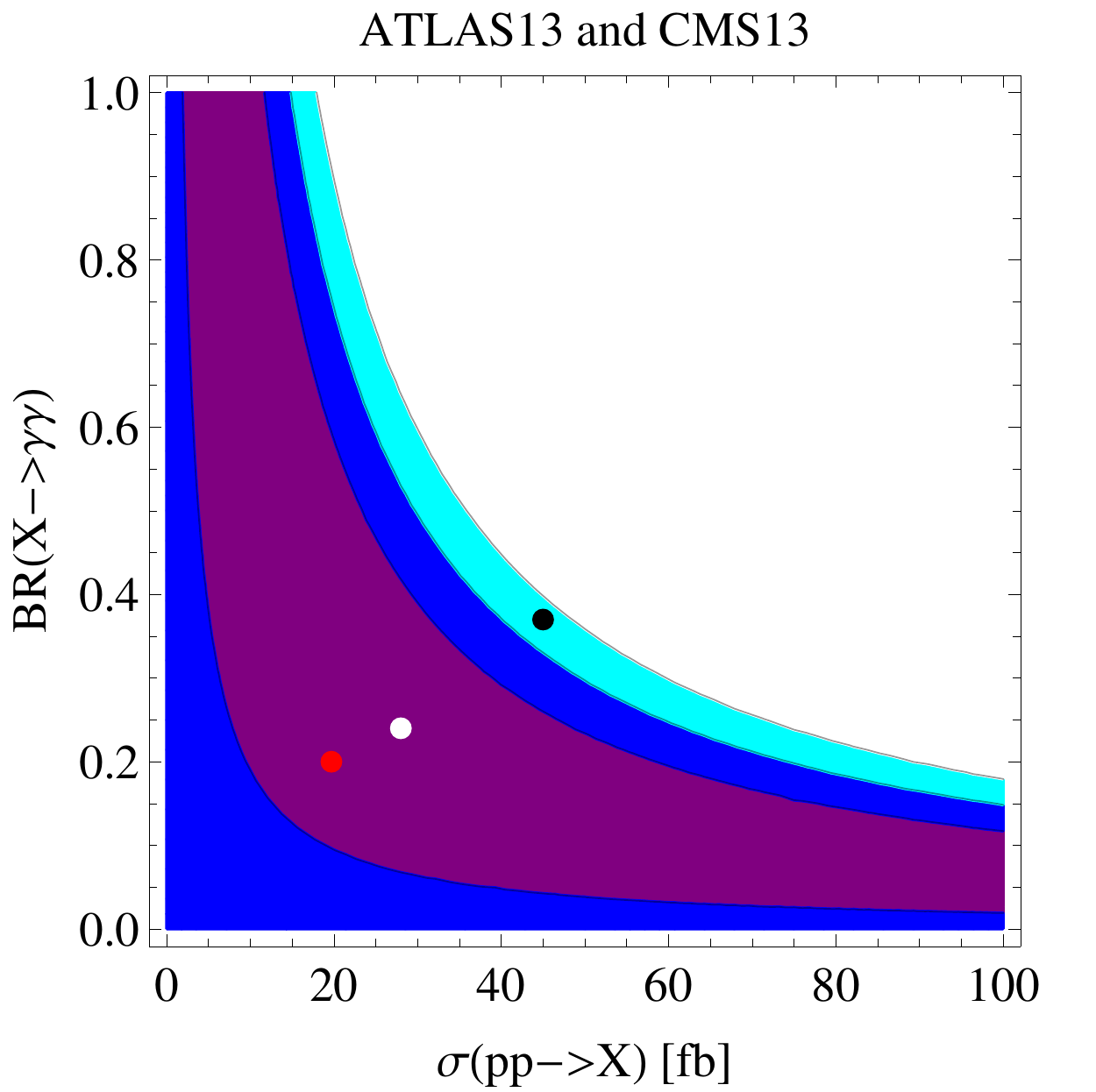} \qquad
\includegraphics[width=0.45\textwidth]{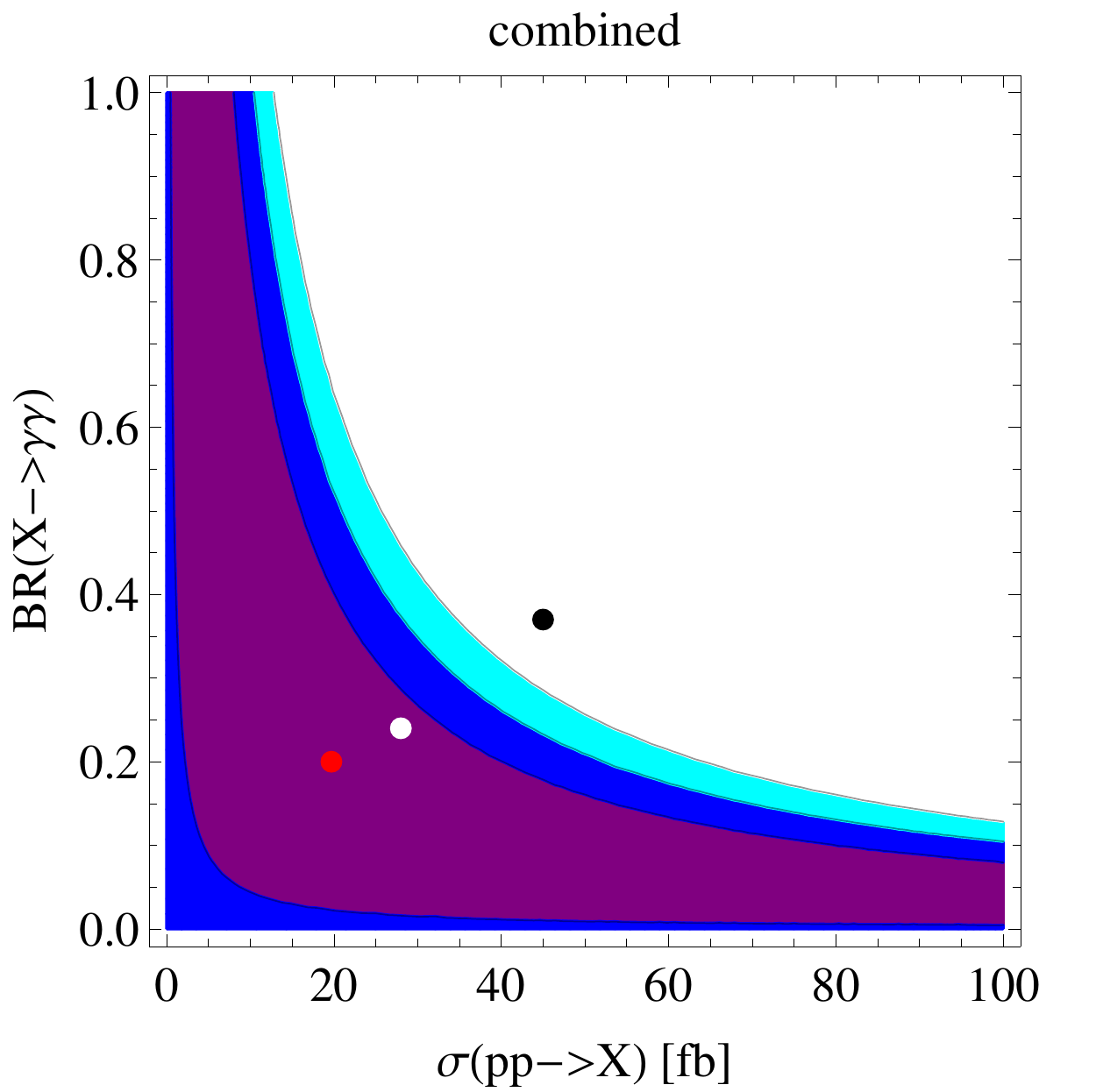} \\
\vspace{-0.75cm}
a) \hspace{0.45\textwidth} b)\hspace{0.45\textwidth}
\caption{The distribution of the $\chi^2$ test as a function of BR$(X\rightarrow\gamma\gamma)$ and $\sigma(pp \to X)$, where $X$ is a spin-2 particle, for a) combined measurements at 13~TeV ATLAS13~\cite{ATLASdiphoton2015} and CMS13~\cite{CMSdiphoton2015} b) combined measurements at 8 and 13~TeV: ATLAS8 EXO~\cite{Aad:2015mna}, CMS8 EXO~\cite{CMS-PAS-EXO-12-045}, ATLAS13~\cite{ATLASdiphoton2015} and CMS13~\cite{CMSdiphoton2015}.  }
\label{fig:summary_2}
\end{center}
\end{figure}

The results are shown in the $\sigma(pp\to X)$-BR$(X \to \gamma\gamma$) plane in Figures~\ref{fig:atlas_2}--\ref{fig:summary_2} for the spin-2 resonance. For each point in the cross section and branching ratio plane, we determined the simulated acceptances for all signal regions and we calculated the $\chi^2$ value as given in Eq.~\eqref{eq:chi2}. Note that for $8$ TeV searches the horizontal scale is the same as for $13$~TeV but in the calculation of the $\chi^2$ and the respective event numbers, the rescaled value is used with the correction factor of $4.7$ due to a gluon luminosity ratio between $8$ and $13$~TeV for a $750$~GeV particle~\cite{Higgs:Cross}. In all plots, different colors denote different level of agreement: purple for 1-$\sigma$ compatibility, blue for 2-$\sigma$, and cyan for 3-$\sigma$. In each plot we also show example points that minimize the $\chi^2$ function: the black point corresponds to the fit using only ATLAS13 data, the white point using both $13$~TeV results and the red point for a combination of $8$ and $13$ 
TeV data. However, one should keep in mind that in each case we obtain a 
hyperbolic line that minimizes the $\chi^2$.

In Figure~\ref{fig:atlas_2}(a) we show the result of the fit using the ATLAS13 data set. The black point denotes one of the good solutions, with $\sigma=45 \fb$ and $\mathrm{BR(X \rightarrow\gamma\gamma)}=37\%$. This can be compared to the remaining measurements. In case of the ATLAS8 HIG search, 
Figure~\ref{fig:atlas_2}(b), we see that this point lies just outside the 2-$\sigma$ contour. When moving to the ATLAS8 EXO search, Figure~\ref{fig:atlas_2}(c), 
there is a clear tension above 3-$\sigma$ as was already seen in Table~\ref{tab:result-spin2}. A similar comparison is made for CMS measurements in 
Figure~\ref{fig:cms_2}. Interestingly, it appears that there is also a significant tension between the ATLAS13 and CMS13 measurements. 
The $8$~TeV CMS searches show compatibility at the 2-$\sigma$ level.

Finally, in Figure~\ref{fig:summary_2}(a) we present a fit to both $13$~TeV results. 
Again, the best-fit point for ATLAS lies between the 2- and 3-$\sigma$ contours. In Figure~\ref{fig:summary_2}(b) 
one can see the results of the fit to both $8$ and $13$~TeV data sets and it is again clearly visible that is difficult to accommodate the ATLAS result point with other measurements. However, we want to stress that for the best fit solution of both 13~TeV results (white point) as well as for the best-fit solution for the combination of 8 and 13~TeV diphoton searches (red point), cf.\ Eq.~\eqref{eq:chi2}, the compatibility is within the 1-$\sigma$ band.
Although we considered here only gluon initiated production, the case for quark initiated processes would result in an even bigger tension between $8$ and $13$ TeV data, since the luminosity ratio for the $q\bar{q}$ initial state is $2.7$~\cite{Higgs:Cross}. 

\begin{figure}[t!]
\begin{center}                                                                                                                                  
\includegraphics[width=0.32\textwidth]{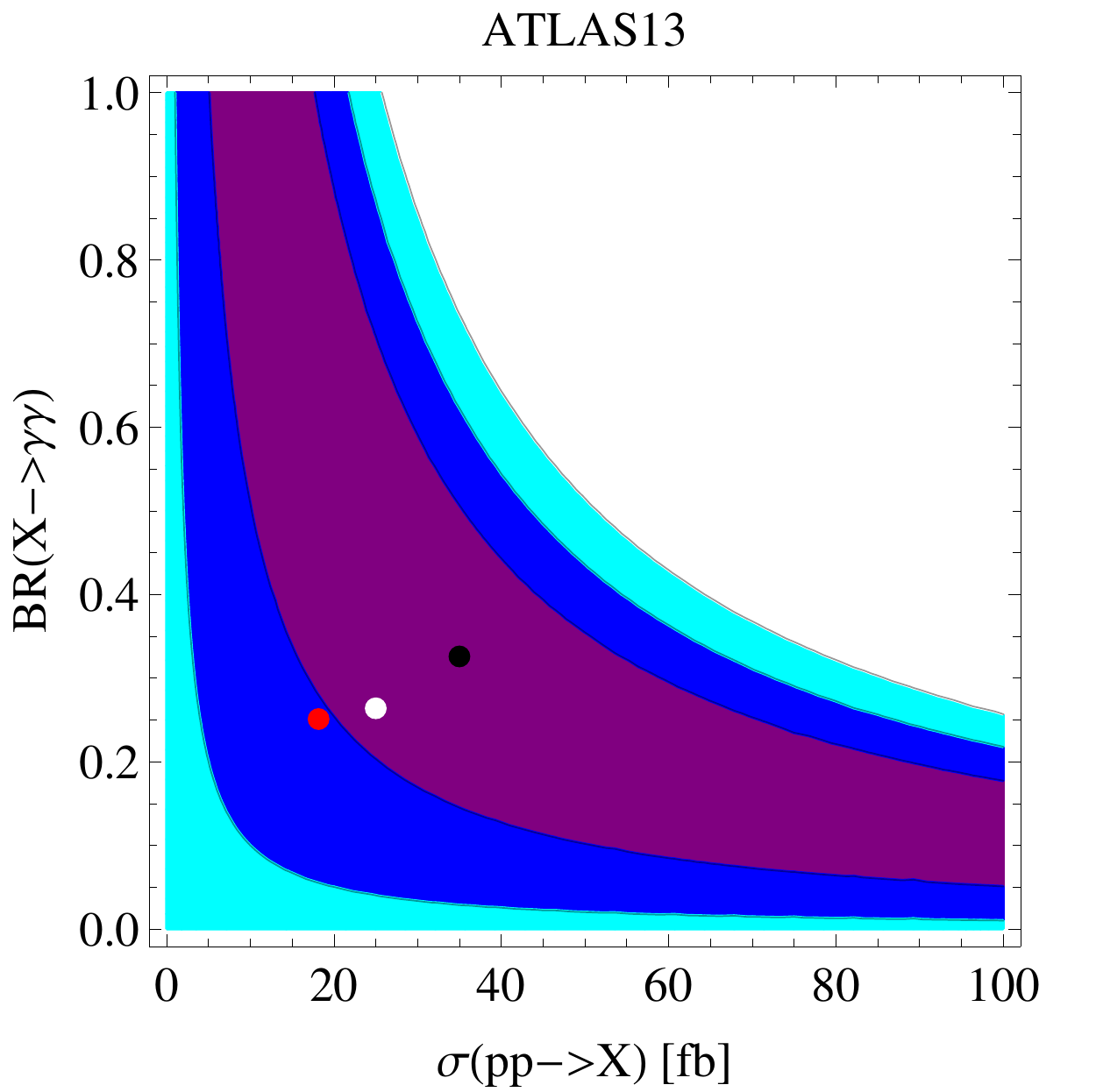} 
\includegraphics[width=0.32\textwidth]{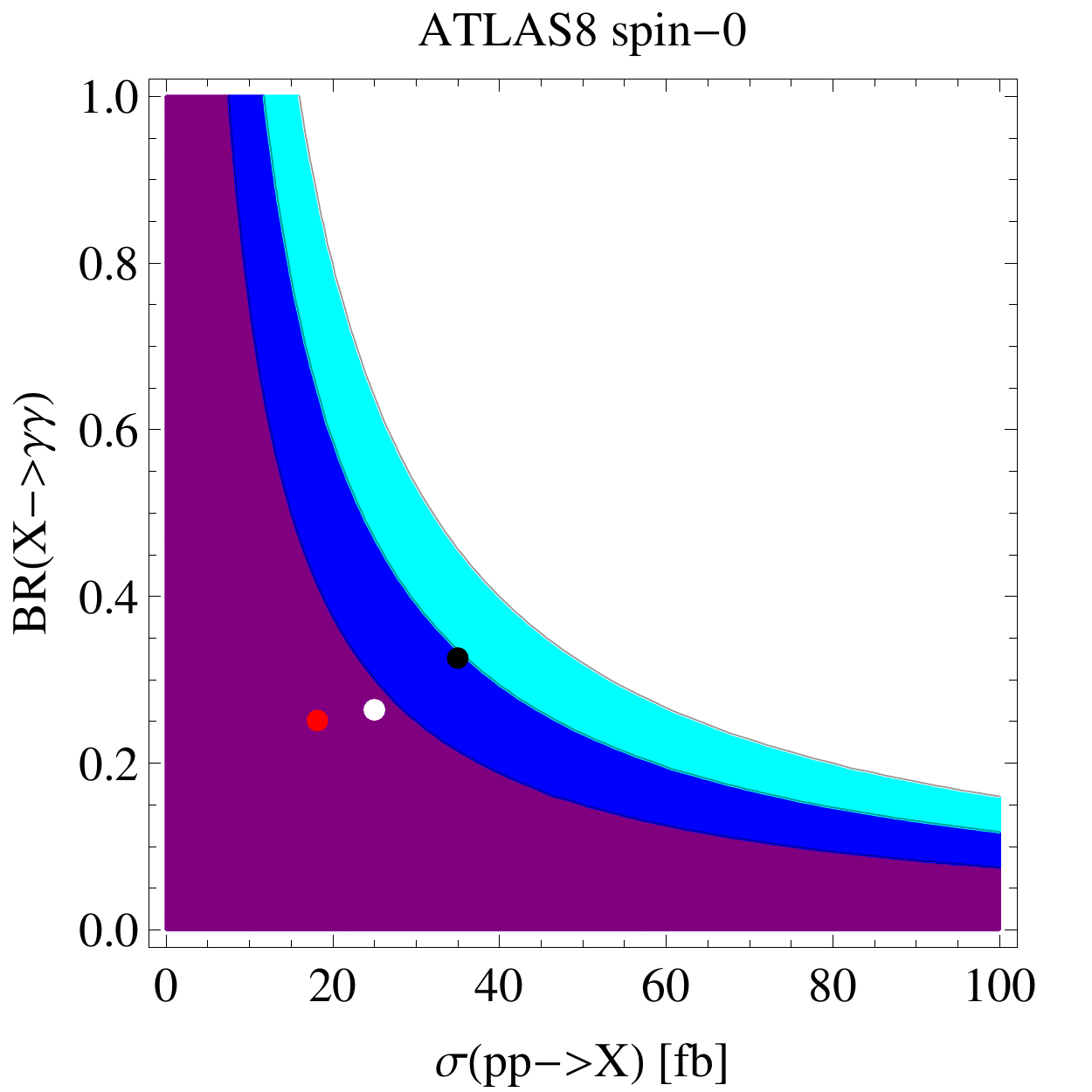}
\includegraphics[width=0.32\textwidth]{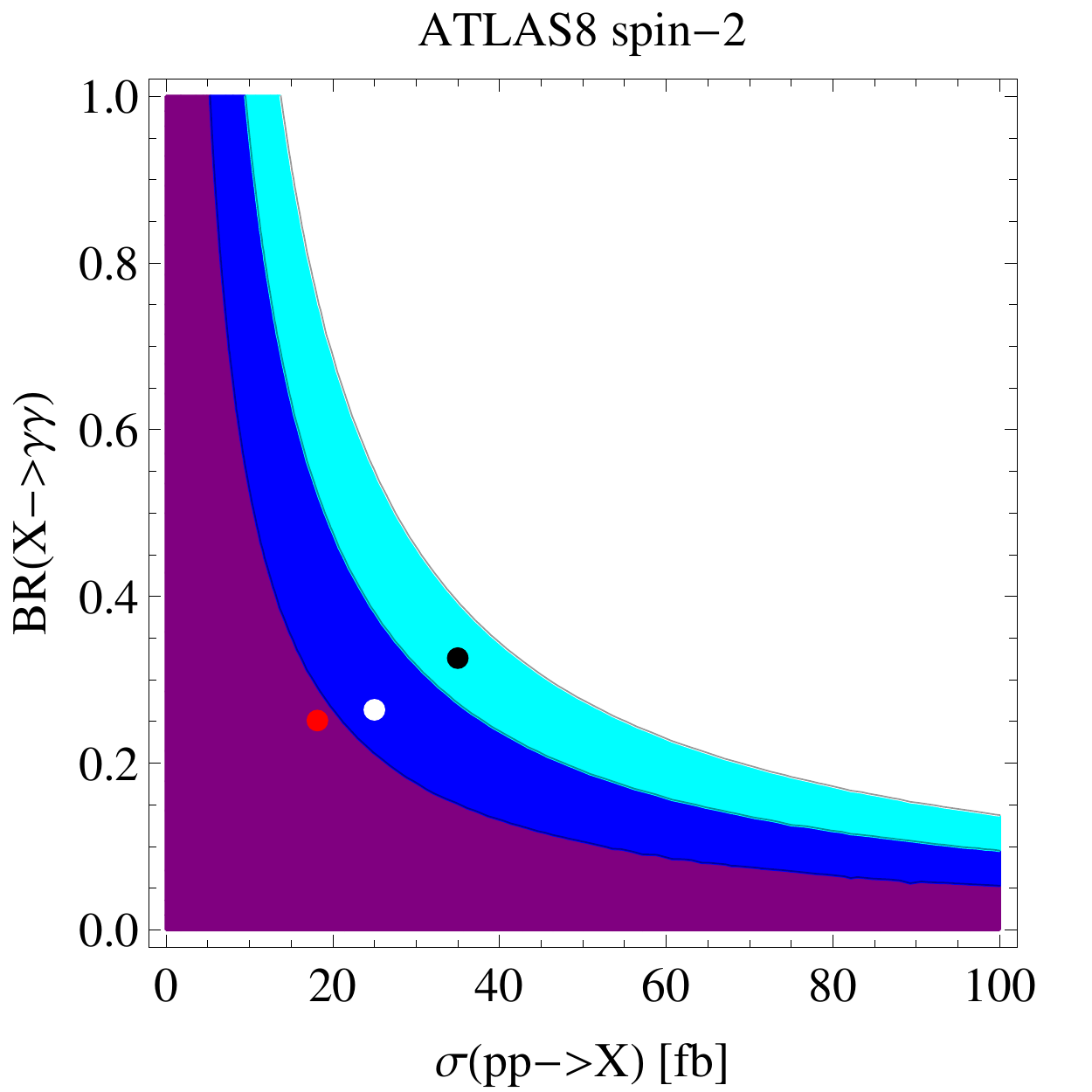}\\
\hspace{0.05\textwidth}(a) \hspace{0.28\textwidth} (b) \hspace{0.28\textwidth} (c)
\caption{The distribution of the $\chi^2$ test as a function of BR$(X\rightarrow\gamma\gamma)$ and $\sigma(pp \to X)$, where $X$ is a spin-0 particle, for (a) ATLAS13~\cite{ATLASdiphoton2015}; (b) ATLAS8 HIG~\cite{Aad:2014ioa}; and (c) ATLAS8 EXO~\cite{Aad:2015mna}. See text and Figure~\ref{fig:atlas_2} for more details. }
\label{fig:atlas_0}
\end{center}
\end{figure}

\begin{figure}[t!]
\begin{center}
\includegraphics[width=0.32\textwidth]{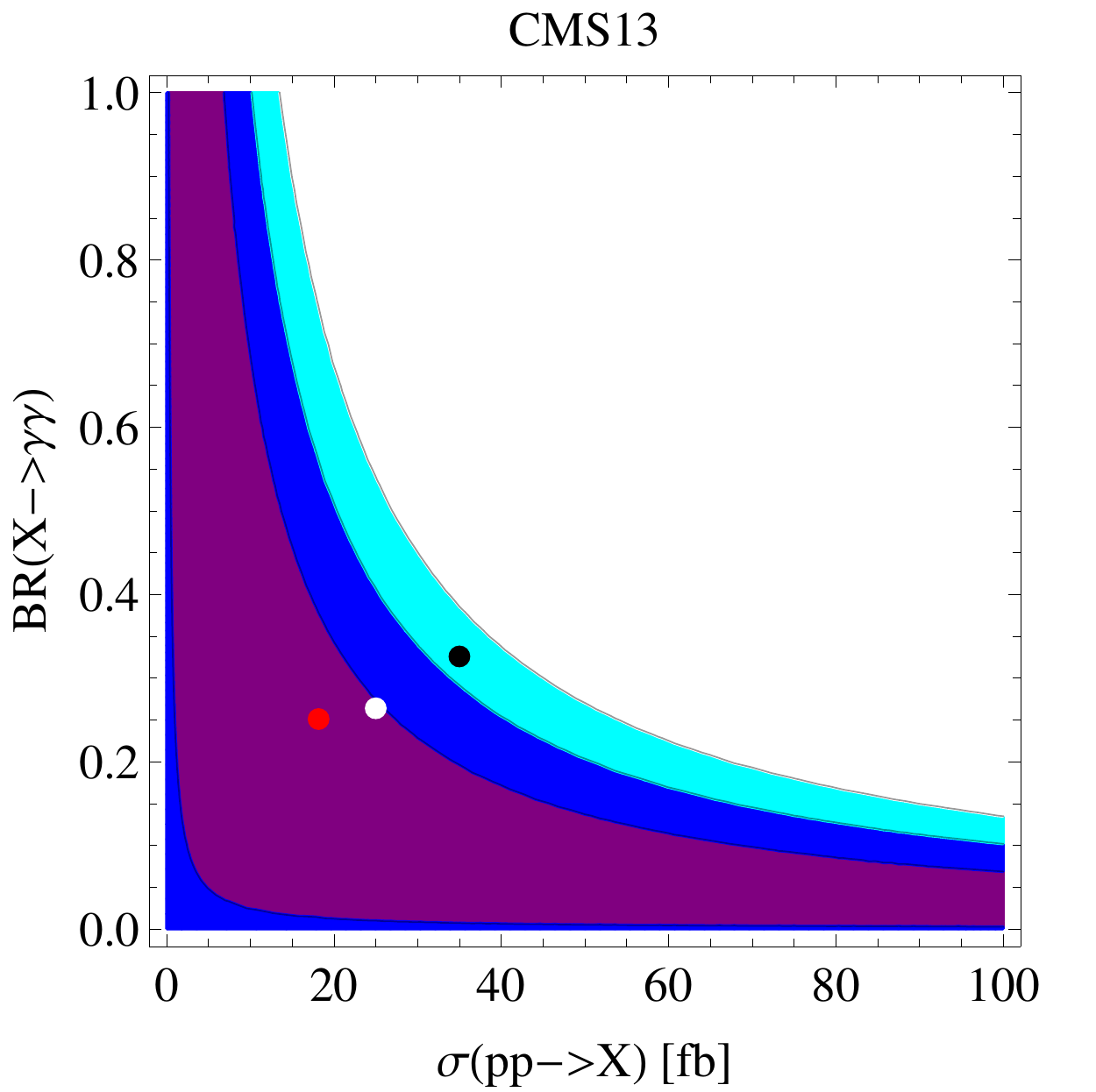} 
\includegraphics[width=0.32\textwidth]{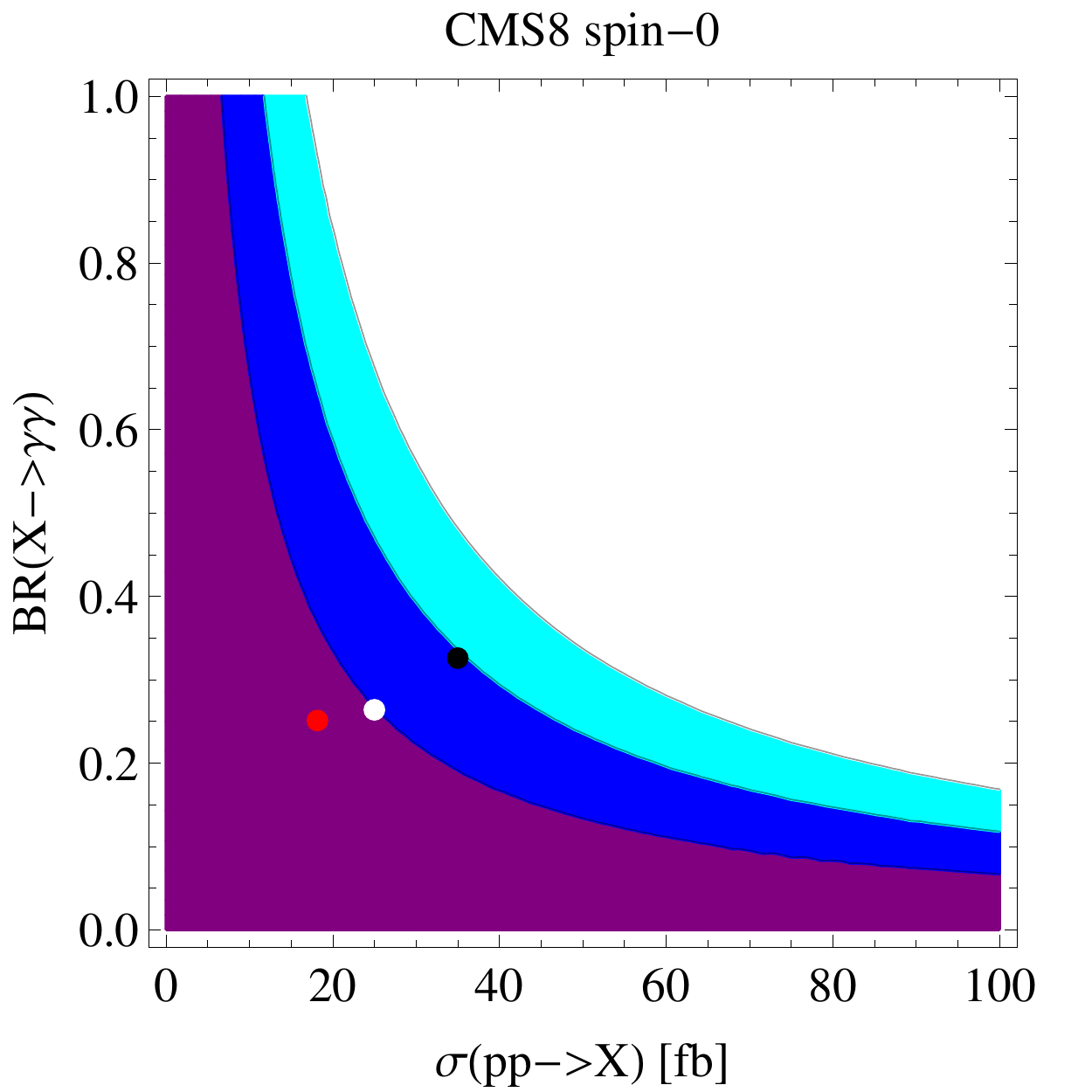} 
\includegraphics[width=0.32\textwidth]{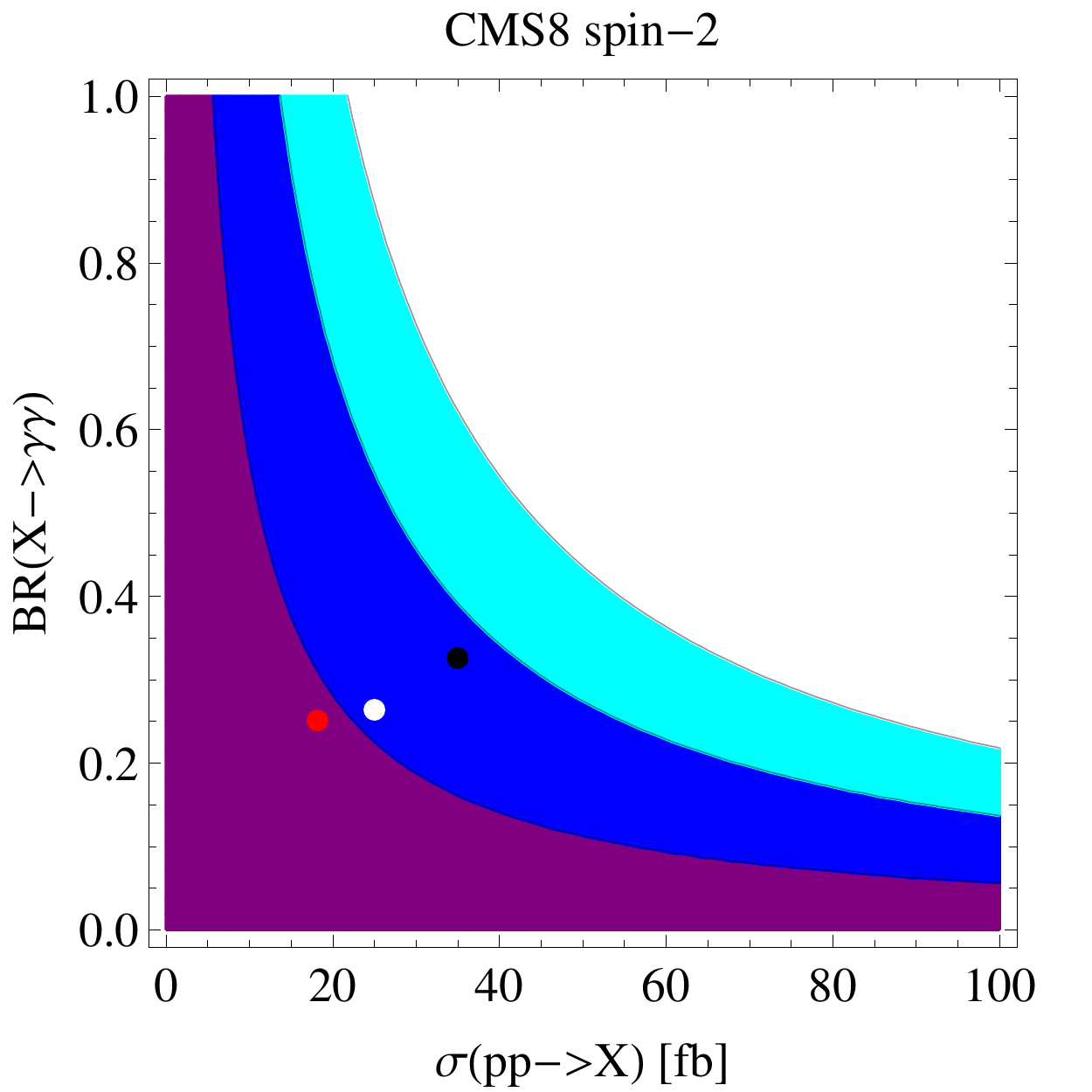} \\
%\vspace{-0.75cm}
\hspace{0.05\textwidth}(a) \hspace{0.28\textwidth} (b) \hspace{0.28\textwidth} (c)
\caption{The distribution of the $\chi^2$ test as a function of BR$(X\rightarrow\gamma\gamma)$ and $\sigma(pp \to X)$, where $X$ is a spin-0 particle, for (a) CMS13~\cite{CMSdiphoton2015}; (b) CMS8 HIG~\cite{Khachatryan:2015qba}; and (c) CMS8 EXO~\cite{CMS-PAS-EXO-12-045}. See text and Figure~\ref{fig:atlas_2} for more details. }
\label{fig:cms_0}
\end{center}
\end{figure}

\begin{figure}[t!]
\begin{center}                                                                                                                                  
\includegraphics[width=0.45\textwidth]{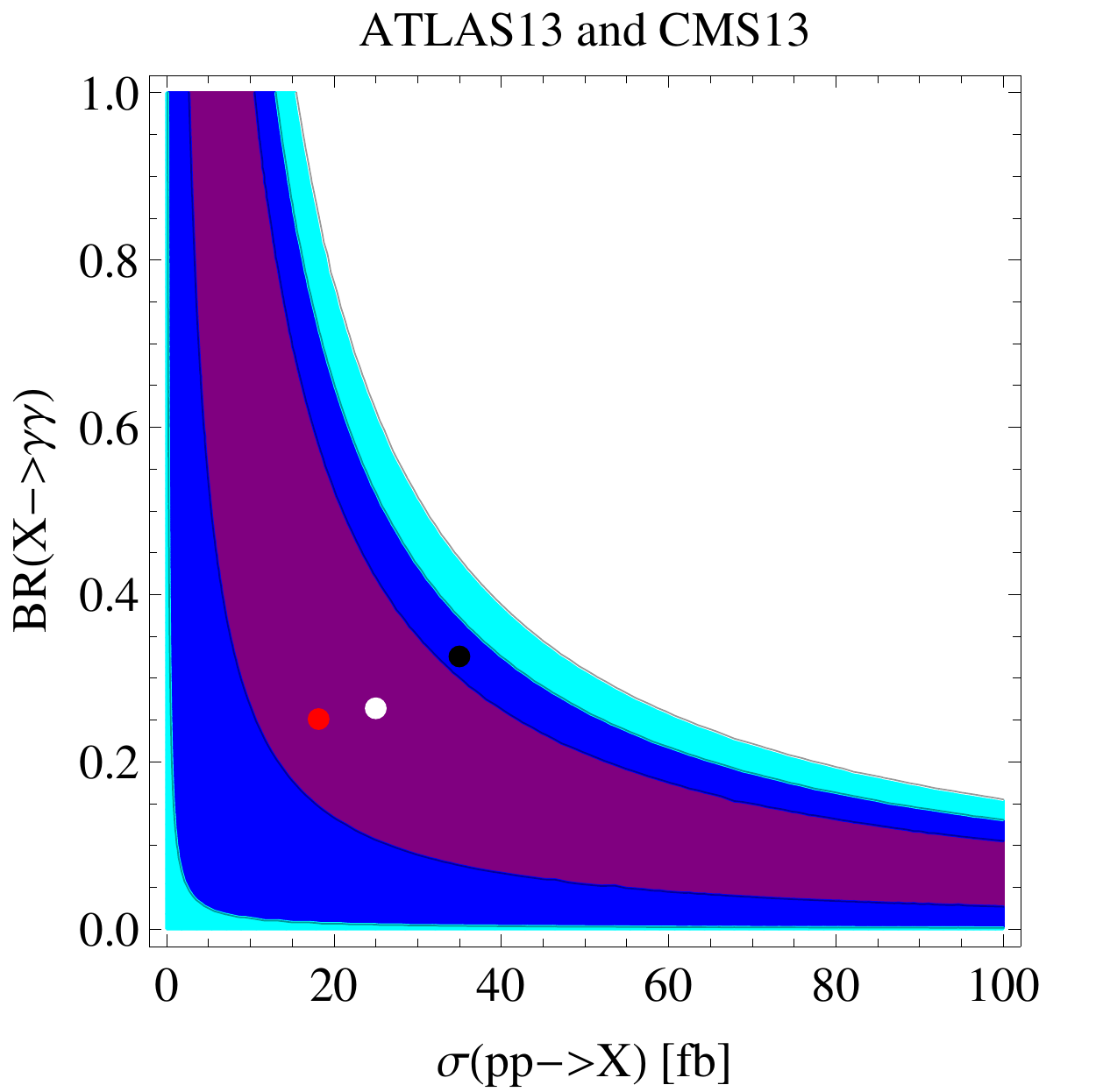} \qquad
\includegraphics[width=0.45\textwidth]{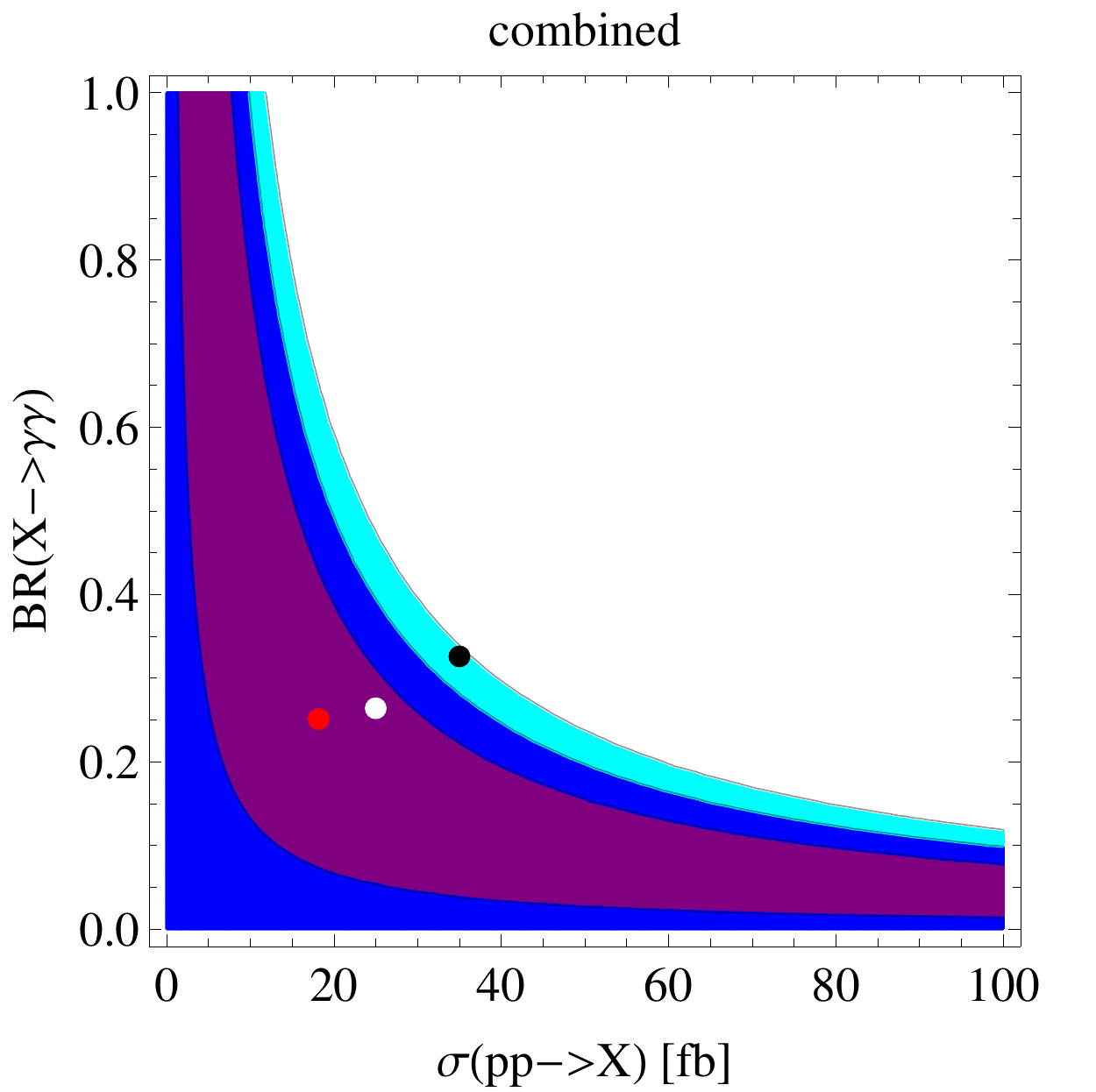} \\
\vspace{-0.75cm}
a) \hspace{0.45\textwidth} b)\hspace{0.45\textwidth}
\caption{The distribution of the $\chi^2$ test as a function of BR$(X\rightarrow\gamma\gamma)$ and $\sigma(pp \to X)$, where $X$ is a spin-0 particle, for a) combined measurements at 13~TeV ATLAS13~\cite{ATLASdiphoton2015} and CMS13~\cite{CMSdiphoton2015} b) combined measurements at 8 and 13~TeV: ATLAS8 EXO~\cite{Aad:2015mna}, CMS8 EXO~\cite{CMS-PAS-EXO-12-045}, ATLAS13~\cite{ATLASdiphoton2015} and CMS13~\cite{CMSdiphoton2015}.   }
\label{fig:summary_0}
\end{center}
\end{figure}

In a similar fashion we also study the production of a spin-0 particle. The results are listed in Table~\ref{tab:result-spin0} and in Figures~\ref{fig:atlas_0}, \ref{fig:cms_0} and \ref{fig:summary_0}. While the results generally follow a similar pattern as for the spin-2 case, the incompatibility between the ATLAS13 measurement and the remaining searches is drastically decreased, e.g.\ the best ATLAS13 fit solution for the spin-0 resonance corresponds to 28.1 signal events in the ATLAS8 EXO search with a $\mathrm{CL}_\mathrm{S}=0.02$. The most significant tension can be seen in Figure~\ref{fig:summary_0}(b) where the ATLAS result just lies below the 3-$\sigma$ contour. 

As already mentioned, there is a flat direction in the $\chi^2$ minimum. This is clear since we have several measurements, however, of the same quantity: $\sigma(p p \to X) \times \mathrm{BR}(X \to \gamma\gamma)$. The degeneracy for the best-fit solution would be lifted by a measurement of the rate of resonance decays into dijet final states.
In the following we provide a functional shape of this minimum for each fit considered. The function that reproduces the minimum is given by
\begin{equation}\label{eq:function}
 \mathrm{BR}(X\to \gamma\gamma) = \frac{a}{\sigma(p p \to X)}\,,
\end{equation}
where $a$ is a free parameter and can be determined from the data. The values of $a$ with uncertainties for different cases are summarized in Table~\ref{tab:avalues}. 

We want to conclude this section with a few comments about other LHC constraints apart from the diphoton searches. In principle, constraints from dijet and diboson searches could also start to play a role at some point. Regarding the dijet searches, the experimental bounds would typically be at the pb level~\cite{Aad:2014aqa}, which is well above the cross sections considered here. Similarly, the diboson searches become relevant for cross sections $\mathcal{O}(100)$~fb~\cite{Aad:2015ipg}. Moreover, those constraints can be avoided depending on exact model assumptions and as a consequence we omit the discussion of other final states.

\begin{table*}[t!]
\begin{center}
\begin{tabular}{|c|c|c|c||c|c|c|}\hline
& \multicolumn{3}{c||}{spin-2} & \multicolumn{3}{c|}{spin-0}\\
\hline
&  ATLAS13 & 13 TeV & combined & ATLAS13 & 13 TeV & combined \\
\hline
$a\ [\mathrm{fb}]$ & $16.7 \pm 6$ & $6.8 \pm 3$ & $4.2 \pm 2.6$ &  $11.4 \pm 7$ & $6.0 \pm 2.5$ & $4.2 \pm 2$ \\
\hline
\end{tabular}
\end{center}
\caption{The values of the $a$ parameter from Eq.~\eqref{eq:function} for a spin-2 and spin-0 resonance and different sets of fitted data: only ATLAS13; both measurements at $13$ TeV; and the combination of $8$ and $13$ TeV data.\label{tab:avalues}}
\end{table*}\label{tab:bestfit}

%%%%%%%%%%%%%%%%%%%%%%%%%%%%%%%%%%
%%%%%%%%%%%%%%%%%%%%%%%%%%%%%%%%%

\section{Conclusions}
In this work, we have tested the compatibility of the diphoton excess between the ATLAS and CMS diphoton searches at the center-of-mass energies of $8$ and $13$~TeV. We considered the resonant production of spin-0 and spin-2 particles. We presented our results in a {\it model-independent} way parametrized in terms of the production cross section and the diphoton branching ratio. The main results of our study are summarized in Tables~\ref{tab:result-spin2} and~\ref{tab:result-spin0} as well as in Figures~\ref{fig:summary_2} and~\ref{fig:summary_0}.

We have found a slight tension between the best-fit solution of the spin-2 scenario for the ATLAS diphoton excess at $13$~TeV and the other diphoton searches. In particular, with the exotic search of ATLAS at $8$~TeV and the CMS result at $13$~TeV where the discrepancy can be larger than 3-$\sigma$ in some cases. 
However, this tension seems to be significantly smaller for the spin-0 hypothesis. 

Finally, we provide a functional form of the relation between the cross section and branching ratio that parametrizes the best fit to the data summarized in Table~\ref{tab:bestfit}. This can be used in future analyses to quickly find whether the relation predicted in a particular model describes the data well.

Clearly, more data at $13$ TeV will be needed to confirm or reject the existence of the excess. We are looking forward to the results of the ongoing Run-II.

%%%%%%%%%%%%%%%%%%%%%%%%%%%%%%%%%%%%%%%%
%%%%%%%%%%%%%%%%%%%%%%%%%%%%%%%%%%%%%%%%

\paragraph{Acknowledgments} 

R. RdA is supported by the Ram\'on y Cajal program of the Spanish
MICINN and also thanks the support of the Spanish MICINN's
Consolider-Ingenio 2010 Programme  under the grant MULTIDARK
CSD2209-00064, the Invisibles European ITN project
(FP7-PEOPLE-2011-ITN, PITN-GA-2011-289442-INVISIBLES) and the  
``SOM Sabor y origen de la Materia" (FPA2011-29678) and the
``Fenomenologia y Cosmologia de la Fisica mas alla del Modelo Estandar
e lmplicaciones Experimentales en la era del LHC" (FPA2010-17747) MEC
projects.  K.R. and J.S.K. has been partially supported by the MINECO (Spain) under contract FPA2013-44773-P; 
Consolider-Ingenio CPAN CSD2007-00042; the Spanish MINECO Centro de excelencia Severo Ochoa 
Program under grant SEV-2012-0249; and by JAE-Doc program. 

%%%%%%%%%%%%%%%%%%%%%%%%%%%%%%%%
%%%%%%%%%%%%%%%%%%%%%%%%%%%%%%%%

\bibliography{resonance}

\providecommand{\href}[2]{#2}\begingroup\raggedright\begin{thebibliography}{10}

\bibitem{ATLASdiphoton2015}
{\bf ATLAS} Collaboration, {\it {Search for resonances decaying to photon pairs
  in 3.2 inverse fb of pp collisions at $\sqrt{s}$ = 13 TeV with the ATLAS
  detector}},  \href{http://arxiv.org/abs/ATLAS-CONF-2015-081}{{\tt
  ATLAS-CONF-2015-081}}.

\bibitem{CMSdiphoton2015}
{\bf CMS} Collaboration, {\it {Search for new physics in high mass diphoton
  events in proton-proton collisions at 13 TeV}},
  \href{http://arxiv.org/abs/CMS PAS EXO-15-004}{{\tt CMS PAS EXO-15-004}}.

\bibitem{Landau:1948kw}
L.~D. Landau, {\it {On the angular momentum of a system of two photons}},  {\em
  Dokl. Akad. Nauk Ser. Fiz.} {\bf 60} (1948), no.~2 207--209.

\bibitem{Yang:1950rg}
C.-N. Yang, {\it {Selection Rules for the Dematerialization of a Particle Into
  Two Photons}},  {\em Phys. Rev.} {\bf 77} (1950) 242--245.

\bibitem{Harigaya:2015ezk}
K.~Harigaya and Y.~Nomura, {\it {Composite Models for the 750 GeV Diphoton
  Excess}},  {\em Phys. Lett.} {\bf B754} (2016) 151--156,
  [\href{http://arxiv.org/abs/1512.04850}{{\tt arXiv:1512.04850}}].

\bibitem{Mambrini:2015wyu}
Y.~Mambrini, G.~Arcadi, and A.~Djouadi, {\it {The LHC diphoton resonance and
  dark matter}},  {\em Phys. Lett.} {\bf B755} (2016) 426--432,
  [\href{http://arxiv.org/abs/1512.04913}{{\tt arXiv:1512.04913}}].

\bibitem{Backovic:2015fnp}
M.~Backovic, A.~Mariotti, and D.~Redigolo, {\it {Di-photon excess illuminates
  Dark Matter}},  {\em JHEP} {\bf 03} (2016) 157,
  [\href{http://arxiv.org/abs/1512.04917}{{\tt arXiv:1512.04917}}].

\bibitem{Angelescu:2015uiz}
A.~Angelescu, A.~Djouadi, and G.~Moreau, {\it {Scenarii for interpretations of
  the LHC diphoton excess: two Higgs doublets and vector-like quarks and
  leptons}},  {\em Phys. Lett.} {\bf B756} (2016) 126--132,
  [\href{http://arxiv.org/abs/1512.04921}{{\tt arXiv:1512.04921}}].

\bibitem{Buttazzo:2015txu}
D.~Buttazzo, A.~Greljo, and D.~Marzocca, {\it {Knocking on new physics' door
  with a scalar resonance}},  {\em Eur. Phys. J.} {\bf C76} (2016), no.~3 116,
  [\href{http://arxiv.org/abs/1512.04929}{{\tt arXiv:1512.04929}}].

\bibitem{Knapen:2015dap}
S.~Knapen, T.~Melia, M.~Papucci, and K.~Zurek, {\it {Rays of light from the
  LHC}},  {\em Phys. Rev.} {\bf D93} (2016), no.~7 075020,
  [\href{http://arxiv.org/abs/1512.04928}{{\tt arXiv:1512.04928}}].

\bibitem{Nakai:2015ptz}
Y.~Nakai, R.~Sato, and K.~Tobioka, {\it {Footprints of New Strong Dynamics via
  Anomaly}},  {\em Phys. Rev. Lett.} {\bf 116} (2016), no.~15 151802,
  [\href{http://arxiv.org/abs/1512.04924}{{\tt arXiv:1512.04924}}].

\bibitem{Pilaftsis:2015ycr}
A.~Pilaftsis, {\it {Diphoton Signatures from Heavy Axion Decays at the CERN
  Large Hadron Collider}},  {\em Phys. Rev.} {\bf D93} (2016), no.~1 015017,
  [\href{http://arxiv.org/abs/1512.04931}{{\tt arXiv:1512.04931}}].

\bibitem{Franceschini:2015kwy}
R.~Franceschini, G.~F. Giudice, J.~F. Kamenik, M.~McCullough, A.~Pomarol,
  R.~Rattazzi, M.~Redi, F.~Riva, A.~Strumia, and R.~Torre, {\it {What is the
  $\gamma \gamma$ resonance at 750 GeV?}},  {\em JHEP} {\bf 03} (2016) 144,
  [\href{http://arxiv.org/abs/1512.04933}{{\tt arXiv:1512.04933}}].

\bibitem{DiChiara:2015vdm}
S.~Di~Chiara, L.~Marzola, and M.~Raidal, {\it {First interpretation of the 750
  GeV di-photon resonance at the LHC}},
  \href{http://arxiv.org/abs/1512.04939}{{\tt arXiv:1512.04939}}.

\bibitem{Higaki:2015jag}
T.~Higaki, K.~S. Jeong, N.~Kitajima, and F.~Takahashi, {\it {The QCD Axion from
  Aligned Axions and Diphoton Excess}},  {\em Phys. Lett.} {\bf B755} (2016)
  13--16, [\href{http://arxiv.org/abs/1512.05295}{{\tt arXiv:1512.05295}}].

\bibitem{McDermott:2015sck}
S.~D. McDermott, P.~Meade, and H.~Ramani, {\it {Singlet Scalar Resonances and
  the Diphoton Excess}},  {\em Phys. Lett.} {\bf B755} (2016) 353--357,
  [\href{http://arxiv.org/abs/1512.05326}{{\tt arXiv:1512.05326}}].

\bibitem{Ellis:2015oso}
J.~Ellis, S.~A.~R. Ellis, J.~Quevillon, V.~Sanz, and T.~You, {\it {On the
  Interpretation of a Possible $\sim 750$ GeV Particle Decaying into $\gamma
  \gamma$}},  {\em JHEP} {\bf 03} (2016) 176,
  [\href{http://arxiv.org/abs/1512.05327}{{\tt arXiv:1512.05327}}].

\bibitem{Low:2015qep}
M.~Low, A.~Tesi, and L.-T. Wang, {\it {A pseudoscalar decaying to photon pairs
  in the early LHC Run 2 data}},  {\em JHEP} {\bf 03} (2016) 108,
  [\href{http://arxiv.org/abs/1512.05328}{{\tt arXiv:1512.05328}}].

\bibitem{Bellazzini:2015nxw}
B.~Bellazzini, R.~Franceschini, F.~Sala, and J.~Serra, {\it {Goldstones in
  Diphotons}},  {\em JHEP} {\bf 04} (2016) 072,
  [\href{http://arxiv.org/abs/1512.05330}{{\tt arXiv:1512.05330}}].

\bibitem{Gupta:2015zzs}
R.~S. Gupta, S.~Jaeger, Y.~Kats, G.~Perez, and E.~Stamou, {\it {Interpreting a
  750 GeV Diphoton Resonance}},  \href{http://arxiv.org/abs/1512.05332}{{\tt
  arXiv:1512.05332}}.

\bibitem{Petersson:2015mkr}
C.~Petersson and R.~Torre, {\it {The 750 GeV diphoton excess from the goldstino
  superpartner}},  {\em Phys. Rev. Lett.} {\bf 116} (2016), no.~15 151804,
  [\href{http://arxiv.org/abs/1512.05333}{{\tt arXiv:1512.05333}}].

\bibitem{Molinaro:2015cwg}
E.~Molinaro, F.~Sannino, and N.~Vignaroli, {\it {Strong dynamics or axion
  origin of the diphoton excess}},  \href{http://arxiv.org/abs/1512.05334}{{\tt
  arXiv:1512.05334}}.

\bibitem{Dutta:2015wqh}
B.~Dutta, Y.~Gao, T.~Ghosh, I.~Gogoladze, and T.~Li, {\it {Interpretation of
  the diphoton excess at CMS and ATLAS}},  {\em Phys. Rev.} {\bf D93} (2016),
  no.~5 055032, [\href{http://arxiv.org/abs/1512.05439}{{\tt
  arXiv:1512.05439}}].

\bibitem{Cao:2015pto}
Q.-H. Cao, Y.~Liu, K.-P. Xie, B.~Yan, and D.-M. Zhang, {\it {A Boost Test of
  Anomalous Diphoton Resonance at the LHC}},
  \href{http://arxiv.org/abs/1512.05542}{{\tt arXiv:1512.05542}}.

\bibitem{Matsuzaki:2015che}
S.~Matsuzaki and K.~Yamawaki, {\it {750 GeV Diphoton Signal from One-Family
  Walking Technipion}},  \href{http://arxiv.org/abs/1512.05564}{{\tt
  arXiv:1512.05564}}.

\bibitem{Kobakhidze:2015ldh}
A.~Kobakhidze, F.~Wang, L.~Wu, J.~M. Yang, and M.~Zhang, {\it {750 GeV diphoton
  resonance in a top and bottom seesaw model}},  {\em Phys. Lett.} {\bf B757}
  (2016) 92--96, [\href{http://arxiv.org/abs/1512.05585}{{\tt
  arXiv:1512.05585}}].

\bibitem{Martinez:2015kmn}
R.~Martinez, F.~Ochoa, and C.~F. Sierra, {\it {Diphoton decay for a $750$ GeV
  scalar dark matter}},  \href{http://arxiv.org/abs/1512.05617}{{\tt
  arXiv:1512.05617}}.

\bibitem{Cox:2015ckc}
P.~Cox, A.~D. Medina, T.~S. Ray, and A.~Spray, {\it {Diphoton Excess at 750 GeV
  from a Radion in the Bulk-Higgs Scenario}},
  \href{http://arxiv.org/abs/1512.05618}{{\tt arXiv:1512.05618}}.

\bibitem{Becirevic:2015fmu}
D.~Be\v{c}irevi\'{c}, E.~Bertuzzo, O.~Sumensari, and R.~Zukanovich~Funchal,
  {\it {Can the new resonance at LHC be a CP-Odd Higgs boson?}},  {\em Phys.
  Lett.} {\bf B757} (2016) 261--267,
  [\href{http://arxiv.org/abs/1512.05623}{{\tt arXiv:1512.05623}}].

\bibitem{No:2015bsn}
J.~M. No, V.~Sanz, and J.~Setford, {\it {See-Saw Composite Higgses at the LHC:
  Linking Naturalness to the $750$ GeV Di-Photon Resonance}},
  \href{http://arxiv.org/abs/1512.05700}{{\tt arXiv:1512.05700}}.

\bibitem{Demidov:2015zqn}
S.~V. Demidov and D.~S. Gorbunov, {\it {On sgoldstino interpretation of the
  diphoton excess}},  \href{http://arxiv.org/abs/1512.05723}{{\tt
  arXiv:1512.05723}}.

\bibitem{Chao:2015ttq}
W.~Chao, R.~Huo, and J.-H. Yu, {\it {The Minimal Scalar-Stealth Top
  Interpretation of the Diphoton Excess}},
  \href{http://arxiv.org/abs/1512.05738}{{\tt arXiv:1512.05738}}.

\bibitem{Fichet:2015vvy}
S.~Fichet, G.~von Gersdorff, and C.~Royon, {\it {Scattering Light by Light at
  750 GeV at the LHC}},  {\em Phys. Rev.} {\bf D93} (2016), no.~7 075031,
  [\href{http://arxiv.org/abs/1512.05751}{{\tt arXiv:1512.05751}}].

\bibitem{Curtin:2015jcv}
D.~Curtin and C.~B. Verhaaren, {\it {Quirky Explanations for the Diphoton
  Excess}},  {\em Phys. Rev.} {\bf D93} (2016), no.~5 055011,
  [\href{http://arxiv.org/abs/1512.05753}{{\tt arXiv:1512.05753}}].

\bibitem{Bian:2015kjt}
L.~Bian, N.~Chen, D.~Liu, and J.~Shu, {\it {A hidden confining world on the 750
  GeV diphoton excess}},  \href{http://arxiv.org/abs/1512.05759}{{\tt
  arXiv:1512.05759}}.

\bibitem{Chakrabortty:2015hff}
J.~Chakrabortty, A.~Choudhury, P.~Ghosh, S.~Mondal, and T.~Srivastava, {\it
  {Di-photon resonance around 750 GeV: shedding light on the theory
  underneath}},  \href{http://arxiv.org/abs/1512.05767}{{\tt
  arXiv:1512.05767}}.

\bibitem{Ahmed:2015uqt}
A.~Ahmed, B.~M. Dillon, B.~Grzadkowski, J.~F. Gunion, and Y.~Jiang, {\it
  {Higgs-radion interpretation of 750 GeV di-photon excess at the LHC}},
  \href{http://arxiv.org/abs/1512.05771}{{\tt arXiv:1512.05771}}.

\bibitem{Agrawal:2015dbf}
P.~Agrawal, J.~Fan, B.~Heidenreich, M.~Reece, and M.~Strassler, {\it
  {Experimental Considerations Motivated by the Diphoton Excess at the LHC}},
  \href{http://arxiv.org/abs/1512.05775}{{\tt arXiv:1512.05775}}.

\bibitem{Csaki:2015vek}
C.~Cs\'{a}ki, J.~Hubisz, and J.~Terning, {\it {Minimal model of a diphoton
  resonance: Production without gluon couplings}},  {\em Phys. Rev.} {\bf D93}
  (2016), no.~3 035002, [\href{http://arxiv.org/abs/1512.05776}{{\tt
  arXiv:1512.05776}}].

\bibitem{Falkowski:2015swt}
A.~Falkowski, O.~Slone, and T.~Volansky, {\it {Phenomenology of a 750 GeV
  Singlet}},  {\em JHEP} {\bf 02} (2016) 152,
  [\href{http://arxiv.org/abs/1512.05777}{{\tt arXiv:1512.05777}}].

\bibitem{Aloni:2015mxa}
D.~Aloni, K.~Blum, A.~Dery, A.~Efrati, and Y.~Nir, {\it {On a possible large
  width 750 GeV diphoton resonance at ATLAS and CMS}},
  \href{http://arxiv.org/abs/1512.05778}{{\tt arXiv:1512.05778}}.

\bibitem{Bai:2015nbs}
Y.~Bai, J.~Berger, and R.~Lu, {\it {A 750 GeV Dark Pion: Cousin of a Dark
  G-parity-odd WIMP}},  {\em Phys. Rev.} {\bf D93} (2016), no.~7 076009,
  [\href{http://arxiv.org/abs/1512.05779}{{\tt arXiv:1512.05779}}].

\bibitem{Gabrielli:2015dhk}
E.~Gabrielli, K.~Kannike, B.~Mele, M.~Raidal, C.~Spethmann, and H.~Veermae,
  {\it {A SUSY Inspired Simplified Model for the 750 GeV Diphoton Excess}},
  {\em Phys. Lett.} {\bf B756} (2016) 36--41,
  [\href{http://arxiv.org/abs/1512.05961}{{\tt arXiv:1512.05961}}].

\bibitem{Benbrik:2015fyz}
R.~Benbrik, C.-H. Chen, and T.~Nomura, {\it {Higgs singlet boson as a diphoton
  resonance in a vectorlike quark model}},  {\em Phys. Rev.} {\bf D93} (2016),
  no.~5 055034, [\href{http://arxiv.org/abs/1512.06028}{{\tt
  arXiv:1512.06028}}].

\bibitem{Kim:2015ron}
J.~S. Kim, J.~Reuter, K.~Rolbiecki, and R.~Ruiz~de Austri, {\it {A resonance
  without resonance: scrutinizing the diphoton excess at 750 GeV}},  {\em Phys.
  Lett.} {\bf B755} (2016) 403--408,
  [\href{http://arxiv.org/abs/1512.06083}{{\tt arXiv:1512.06083}}].

\bibitem{Alves:2015jgx}
A.~Alves, A.~G. Dias, and K.~Sinha, {\it {The 750 GeV $S$-cion: Where else
  should we look for it?}},  {\em Phys. Lett.} {\bf B757} (2016) 39--46,
  [\href{http://arxiv.org/abs/1512.06091}{{\tt arXiv:1512.06091}}].

\bibitem{Megias:2015ory}
E.~Megias, O.~Pujolas, and M.~Quiros, {\it {On dilatons and the LHC diphoton
  excess}},  \href{http://arxiv.org/abs/1512.06106}{{\tt arXiv:1512.06106}}.

\bibitem{Carpenter:2015ucu}
L.~M. Carpenter, R.~Colburn, and J.~Goodman, {\it {Supersoft SUSY Models and
  the 750 GeV Diphoton Excess, Beyond Effective Operators}},
  \href{http://arxiv.org/abs/1512.06107}{{\tt arXiv:1512.06107}}.

\bibitem{Bernon:2015abk}
J.~Bernon and C.~Smith, {\it {Could the width of the diphoton anomaly signal a
  three-body decay?}},  {\em Phys. Lett.} {\bf B757} (2016) 148--153,
  [\href{http://arxiv.org/abs/1512.06113}{{\tt arXiv:1512.06113}}].

\bibitem{Buckley:2016mbr}
M.~R. Buckley, {\it {Wide or Narrow? The Phenomenology of 750 GeV Diphotons}},
  \href{http://arxiv.org/abs/1601.04751}{{\tt arXiv:1601.04751}}.

\bibitem{Staub:2016dxq}
F.~Staub et~al., {\it {Precision tools and models to narrow in on the 750 GeV
  diphoton resonance}},  \href{http://arxiv.org/abs/1602.05581}{{\tt
  arXiv:1602.05581}}.

\bibitem{Aad:2014ioa}
{\bf ATLAS} Collaboration, G.~Aad et~al., {\it {Search for Scalar Diphoton
  Resonances in the Mass Range $65-600$ GeV with the ATLAS Detector in $pp$
  Collision Data at $\sqrt{s}$ = 8 $TeV$}},  {\em Phys. Rev. Lett.} {\bf 113}
  (2014), no.~17 171801, [\href{http://arxiv.org/abs/1407.6583}{{\tt
  arXiv:1407.6583}}].

\bibitem{Aad:2015mna}
{\bf ATLAS} Collaboration, G.~Aad et~al., {\it {Search for high-mass diphoton
  resonances in $pp$ collisions at $\sqrt{s}=8$ TeV with the ATLAS detector}},
  {\em Phys. Rev.} {\bf D92} (2015), no.~3 032004,
  [\href{http://arxiv.org/abs/1504.05511}{{\tt arXiv:1504.05511}}].

\bibitem{Khachatryan:2015qba}
{\bf CMS} Collaboration, V.~Khachatryan et~al., {\it {Search for diphoton
  resonances in the mass range from 150 to 850 GeV in pp collisions at
  $\sqrt{s} =$ 8 TeV}},  {\em Phys. Lett.} {\bf B750} (2015) 494--519,
  [\href{http://arxiv.org/abs/1506.02301}{{\tt arXiv:1506.02301}}].

\bibitem{Barger:1987nn}
V.~D. Barger and R.~J.~N. Phillips, {\em {COLLIDER PHYSICS}}.
\newblock 1987.

\bibitem{Higgs:Cross}
L.~H. C. S.~W. Group, {\it {LHC Higgs Cross Section Working Group}},
  \href{http://arxiv.org/abs/https://twiki.cern.ch/twiki/bin/view/LHCPhysics/CrossSections}{{\tt
  https://twiki.cern.ch/twiki/bin/view/LHCPhysics/CrossSections}}.

\bibitem{Domingo:2016unq}
F.~Domingo, S.~Heinemeyer, J.~S. Kim, and K.~Rolbiecki, {\it {The NMSSM lives -
  with the 750 GeV diphoton excess}},
  \href{http://arxiv.org/abs/1602.07691}{{\tt arXiv:1602.07691}}.

\bibitem{Bernon:2016dow}
J.~Bernon, A.~Goudelis, S.~Kraml, K.~Mawatari, and D.~Sengupta, {\it
  {Characterising the 750 GeV diphoton excess}},
  \href{http://arxiv.org/abs/1603.03421}{{\tt arXiv:1603.03421}}.

\bibitem{Nason:2004rx}
P.~Nason, {\it {A New method for combining NLO QCD with shower Monte Carlo
  algorithms}},  {\em JHEP} {\bf 11} (2004) 040,
  [\href{http://arxiv.org/abs/hep-ph/0409146}{{\tt hep-ph/0409146}}].

\bibitem{Frixione:2007vw}
S.~Frixione, P.~Nason, and C.~Oleari, {\it {Matching NLO QCD computations with
  Parton Shower simulations: the POWHEG method}},  {\em JHEP} {\bf 11} (2007)
  070, [\href{http://arxiv.org/abs/0709.2092}{{\tt arXiv:0709.2092}}].

\bibitem{Alioli:2010xd}
S.~Alioli, P.~Nason, C.~Oleari, and E.~Re, {\it {A general framework for
  implementing NLO calculations in shower Monte Carlo programs: the POWHEG
  BOX}},  {\em JHEP} {\bf 06} (2010) 043,
  [\href{http://arxiv.org/abs/1002.2581}{{\tt arXiv:1002.2581}}].

\bibitem{Sjostrand:2006za}
T.~Sjostrand, S.~Mrenna, and P.~Z. Skands, {\it {PYTHIA 6.4 Physics and
  Manual}},  {\em JHEP} {\bf 05} (2006) 026,
  [\href{http://arxiv.org/abs/hep-ph/0603175}{{\tt hep-ph/0603175}}].

\bibitem{Lai:2010vv}
H.-L. Lai, M.~Guzzi, J.~Huston, Z.~Li, P.~M. Nadolsky, J.~Pumplin, and C.~P.
  Yuan, {\it {New parton distributions for collider physics}},  {\em Phys.
  Rev.} {\bf D82} (2010) 074024, [\href{http://arxiv.org/abs/1007.2241}{{\tt
  arXiv:1007.2241}}].

\bibitem{Bahr:2008pv}
M.~Bahr et~al., {\it {Herwig++ Physics and Manual}},  {\em Eur. Phys. J.} {\bf
  C58} (2008) 639--707, [\href{http://arxiv.org/abs/0803.0883}{{\tt
  arXiv:0803.0883}}].

\bibitem{Martin:2009iq}
A.~D. Martin, W.~J. Stirling, R.~S. Thorne, and G.~Watt, {\it {Parton
  distributions for the LHC}},  {\em Eur. Phys. J.} {\bf C63} (2009) 189--285,
  [\href{http://arxiv.org/abs/0901.0002}{{\tt arXiv:0901.0002}}].

\bibitem{CMS-PAS-EXO-12-045}
{\bf CMS} Collaboration, {\it {Search for High-Mass Diphoton Resonances in pp
  Collisions at $\sqrt{s}=8$ TeV with the CMS Detector}},  Tech. Rep.
  CMS-PAS-EXO-12-045, CERN, Geneva, 2015.

\bibitem{Drees:2013wra}
M.~Drees, H.~Dreiner, D.~Schmeier, J.~Tattersall, and J.~S. Kim, {\it
  {CheckMATE: Confronting your Favourite New Physics Model with LHC Data}},
  {\em Comput. Phys. Commun.} {\bf 187} (2014) 227--265,
  [\href{http://arxiv.org/abs/1312.2591}{{\tt arXiv:1312.2591}}].

\bibitem{Kim:2015wza}
J.~S. Kim, D.~Schmeier, J.~Tattersall, and K.~Rolbiecki, {\it {A framework to
  create customised LHC analyses within CheckMATE}},  {\em Comput. Phys.
  Commun.} {\bf 196} (2015) 535--562,
  [\href{http://arxiv.org/abs/1503.01123}{{\tt arXiv:1503.01123}}].

\bibitem{deFavereau:2013fsa}
{\bf DELPHES 3} Collaboration, J.~de~Favereau, C.~Delaere, P.~Demin,
  A.~Giammanco, V.~Lema\^{i}tre, A.~Mertens, and M.~Selvaggi, {\it {DELPHES 3,
  A modular framework for fast simulation of a generic collider experiment}},
  {\em JHEP} {\bf 02} (2014) 057, [\href{http://arxiv.org/abs/1307.6346}{{\tt
  arXiv:1307.6346}}].

\bibitem{Read:2002hq}
A.~L. Read, {\it {Presentation of search results: The CL(s) technique}},  {\em
  J. Phys.} {\bf G28} (2002) 2693--2704. [,11(2002)].

\bibitem{Aad:2014aqa}
{\bf ATLAS} Collaboration, G.~Aad et~al., {\it {Search for new phenomena in the
  dijet mass distribution using $p-p$ collision data at $\sqrt{s}=8$ TeV with
  the ATLAS detector}},  {\em Phys. Rev.} {\bf D91} (2015), no.~5 052007,
  [\href{http://arxiv.org/abs/1407.1376}{{\tt arXiv:1407.1376}}].

\bibitem{Aad:2015ipg}
{\bf ATLAS} Collaboration, G.~Aad et~al., {\it {Combination of searches for
  $WW$, $WZ$, and $ZZ$ resonances in $pp$ collisions at $\sqrt{s} = 8$ TeV with
  the ATLAS detector}},  {\em Phys. Lett.} {\bf B755} (2016) 285--305,
  [\href{http://arxiv.org/abs/1512.05099}{{\tt arXiv:1512.05099}}].

\end{thebibliography}\endgroup
\bibliographystyle{JHEP}

\end{document}